\documentclass[aps,prr,superscriptaddress,twocolumn]{revtex4-1}

\usepackage{graphicx}
\usepackage[utf8]{inputenc}
\usepackage[english]{babel}
\usepackage{amsmath,graphicx,enumerate}
\usepackage{epsfig}
\usepackage{hyperref}
\usepackage{amssymb}

\usepackage{color}

\DeclareRobustCommand{\vect}[1]{
  \ifcat#1\relax
    \boldsymbol{#1}
  \else
    \mathbf{#1}
  \fi}

\begin{document}
\title{Understanding causation via correlations and linear response theory}

\author{Marco Baldovin}
\affiliation{Dipartimento di Fisica, Universit\`a di Roma Sapienza, P.le Aldo Moro 5, 00185, Rome, Italy}
\author{Fabio Cecconi}
\affiliation{Istituto dei Sistemi Complessi - CNR, Via dei Taurini 19, 00185, Rome, Italy}
\author{Angelo Vulpiani}
\affiliation{Dipartimento di Fisica, Universit\`a di Roma Sapienza, P.le Aldo Moro 5, 00185, Rome, Italy}

\begin{abstract}
  In spite of the (correct) common-wisdom statement {\it correlation does not 
imply causation}, a proper employ of  time correlations and of 
fluctuation-response theory allows to understand the causal relations between 
the variables of a multi-dimensional linear Markov process. It is shown that the 
fluctuation-response formalism can be used both to find the direct causal links 
between the variables of a system and to introduce a degree of causation, 
cumulative in time, whose physical interpretation is straightforward. Although 
for generic non-linear dynamics there is no simple exact relationship between 
correlations and response functions, the described protocol can still give a 
useful proxy also in presence of weak nonlinear terms.
\end{abstract}

\maketitle

\section{Introduction}
Detection of causation is a fundamental topic in science, whose origin dates 
back to the philosophical investigation of D. Hume~\cite{Hume} and to the roots 
of physical thinking. In its most general terms, the problem may be formulated 
as follows: given the time series $\{x^{(1)}_t\}, \{x^{(2)}_t\},\ldots, 
\{x^{(n)}_t\}$ of $n$ variables constituting an observable system $\vect{x}_t$, 
one wishes to determine unambiguously whether the behavior of $x^{(k)}$ has been 
influenced by $x^{(j)}$ during the dynamics, without knowing the underlying 
evolution laws. Causal detection has a primary practical relevance in physical 
modeling~\cite{PearlBook,Aurell2016causal,zeng20}, where the problem of 
inferring models from data is typically 
faced~\cite{friedrich11,zeng11,baldovin18,ferretti19}. A natural idea, 
summarized by the Latin saying {\it cum hoc ergo propter hoc} (``with this, 
therefore because of this''), is looking at the correlation 
$C_{jk}(t)=\langle x^{(j)}_t x^{(k)}_0\rangle$, since a causal link should lead to a non-zero value 
for it, at least for some $t>0$. On the other hand, the presence of correlation 
does not imply causation, as it is possible, for instance, that both  $x^{(k)}$ 
and $x^{(j)}$ are influenced by one or more common-causal 
variables~\cite{PearlBook,simon54,CausaInfoTh,atmanspacher19}.

A more reliable way to detect the presence of causal effects between two 
variables is the popular Granger causality (GC) test~\cite{Granger69}. This 
method allows to determine whether the knowledge of the past history of 
$x^{(j)}$  enhances the ability to predict future values of $x^{(k)}$. 
Basically, it compares the statistical uncertainties of two predictions 
built on 
the linear regression of past data, obtained by including or ignoring the 
trajectory of $x^{(j)}$. The improvement of the prediction, defined by 
the relative reduction of the uncertainty, gives a measure of how much $x^{(j)}$ is useful to 
the determination of  $x^{(k)}$~\cite{bressler11,barrett10,cadotte08}. A similar approach consists in defining 
a degree of information exchange from $x^{(j)}$ to $x^{(k)}$, which quantifies 
the loss of information about $x^{(k)}$ that one experiences if $\{x^{(j)}_t\}$ 
is ignored. This is exactly what is done by transfer entropy (TE) and related 
quantities~\cite{shreiber00,bossomaier,runge12,runge12pre,sun15} (which also 
have interesting interpretations in the context of information 
thermodynamics~\cite{ito13,auconi19}). Remarkably, TE has been shown to be 
exactly equivalent to GC in linear autoregressive 
systems~\cite{barnett09,barrett10,Equivalence}.

Even if GC, TE and similar quantities can provide useful information about the 
dynamics, their employment as a measure of causal relations may be not 
completely satisfactory from a physical point of view. Indeed, in physics two 
variables are usually believed to be in a cause-effect relationship if an 
external action on one of them results in a change of the observed value of the 
second~\cite{barnett09,Aurell2016causal}, whereas the above mentioned tests, strictly speaking, only 
determine whether, ant to what extent, the knowledge of a certain variable is 
useful to the actual determination of future values of another. In the 
following, we will call ``interventional'' the former, physics-inspired definition of cause-effect 
relation and ``observational'' the latter. Sometimes a similar distinction is 
made between the two approaches, distinguishing between the detection of
``causal mechanisms'' and ``causal effects''~\cite{barrett13,barnett18}. As we will 
discuss in the next Section, the strength of the interventional causal link is 
quantified by a well-known observable, the physical response~\cite{marconi08}, 
whose usage to infer causal relations from data is the main subject of this 
paper.

To clarify the above distinction between interventional and observational 
causation, let us briefly discuss a simple situation in which this difference 
may be relevant. Imagine that we want to measure the electrical current passing 
trough a resistor, when its extremities are connected to an external 
time-dependent source of electric potential, $v(t)$. Let us assume that the 
amperometer we are using is affected by some noise $\eta(t)$ independent of 
$v(t)$. In this case, the measured value of the current $j(t)$ is given by
\begin{equation}
 j_{meas}(t) = j_{true}(t)+\eta(t) =  Gv(t) + \eta(t)\,,
\end{equation} 
where $j_{true}$ is the actual (unknown) value of the current and $G$ is the 
electrical conductance of the considered resistor. In this case, a good 
estimator of the interventional causality between $v(t)$ and $j_{meas}(t)$ will 
only depend on the conductance $G$, since this parameter establishes to which 
extent an external action on $v(t)$ will influence the observed value of the 
current, $j_{meas}(t)$ (a notion which does not depend on the intensity of the 
noise). Conversely, from an observational perspective also the amplitude of the 
noise $\eta(t)$ does play a role, since our ability to predict future values of 
$j_{meas}$, given $v(t)$, crucially depends on it: roughly speaking, if the 
noise is small, the knowledge of $v(t)$ will suffice to give a good esteem of 
$j_{meas}(t)$, whereas if it is large, the information about $v(t)$ is almost 
useless.

In this paper we show that linear response theory allows to understand causal 
links (in the interventional sense) from time series of data, if the 
considered process is of Markov type. Moreover if the dynamics is also linear, 
only simple time correlation functions have to be taken into account. 
This approach 
can be used both to quantify the overall influence of $x^{(j)}$ on  $x^{(k)}$, 
including the effects due to indirect causation, and to infer the matrix of 
direct links between the elements of the system.

Of course, in most cases an analysis based on linear response will provide 
results qualitatively similar to those obtained by mean of TE or GC, since 
information transfer and physical interaction are usually related; however, the 
analytical forms of TE and GC are typically cumbersome, even for very simple 
models, and this makes very difficult to get any insight into the structure of 
the considered system by mean of these tools. Moreover they are usually 
difficult to apply in practical situations, as in experiments, if the 
dimensionality of the system is not very small. The method presented here is 
instead very simple to apply in practice, and its physical interpretation is 
straightforward; the drawback is its rigorous validity only for Markov systems 
with linear dynamics: generalizations to non-linear evolutions are also 
possible, provided that the stationary joint probability density function of the 
system is known.

The paper is structured as follows. In Section~\ref{sec:definition} we give a physical definition of causation using 
the formalism of linear response theory, which is briefly recalled in 
Appendix~\ref{app:fdr}. Section~\ref{sec:linearmarkov} is devoted to linear 
Markov systems: we discuss how the response formalism can be used to infer 
causal links from correlations, and we outline the main differences with other 
approaches. In Section~\ref{sec:generalization} we consider more general 
cases, i.e. non-linear systems and dynamics with hidden variables, and we 
discuss the limits of causation determination from data. Finally, in 
Section~\ref{sec:conclusions} we draw our conclusions.

\section{A physical definition of causation}
\label{sec:definition}

As mentioned in the Introduction, we are mainly interested in the study of causation
in the interventional sense, i.e. the one accounting for the effects of 
external actions of the system, as in typical experimental setups.
Let us consider the 
system ${\mathbf x}_t=(x^{(1)}_t,x^{(2)}_t,...,x^{(n)}_t)$, where $t$ is a (discrete)
time index. 
We say that $x^{(j)}$ influences $x^{(k)}$
if a perturbation on the variable $x^{(j)}$ at time $0$, 
$x^{(j)}_0 \to x^{(j)}_0 +\delta x^{(j)}_0$, induces, on average, 
a change on $x^{(k)}_{t}$, with $t>0$.
In formulae, we will say that $x^{(j)}$ has an influence on $x^{(k)}$ 
if a smooth function $\mathcal{F}(x)$ exists such that
\begin{equation}
 \label{eq:resp_obs}
 \dfrac{ \overline{\delta \mathcal{F}(x^{(k)}_t)}}{\delta x^{(j)}_0} \ne 0 \quad \quad \quad \mbox{for some }t>0\,,
\end{equation}
i.e. if perturbing $x^{(j)}_0$ results in a non-zero average variation of 
$\mathcal{F}(x^{(k)}_t)$
with respect to its unperturbed evolution. Here the over-line represents an 
average over many realizations of the experiment. 
Since we will mainly deal with linear Markov systems, 
considering the identity function $\mathcal{F}(x)=x$ will be sufficient to detect 
the presence of causal links (see Appendix~\ref{app:linear} for a brief discussion on this point).

This idea is not completely new~\cite{barnett09,Aurell2016causal}, and it is 
reminiscent of the framework developed by Pearl~\cite{PearlBook}, in which 
causation is detected by observing the effects of an action on the system 
(although in that context the role of time is not explicitly considered). In particular,
in Pearls' formalism one has to evaluate conditional probabilities assuming that
the graph of the interactions between variables is actively manipulated. A similar
idea can be found in the ``flow of information'' introduced in Ref.~\cite{ay08}, which can be seen as
the information-theoretic counterpart of the Pearl's probabilistic  formalism.

 If the 
system admits a (sufficiently smooth) invariant distribution, and $\delta 
x^{(j)}_0$ is small enough, quantities of the form~\eqref{eq:resp_obs} can be 
evaluated without actually perturbing the system, since they are related to the 
spontaneous correlations in the unperturbed dynamics by the fluctuation-response 
(FR) theorem~\cite{kubo_response,marconi08}, also known as 
fluctuation-dissipation theorem. If $\{\vect{x}_t\}$ is a stationary process 
with invariant probability density function (p.d.f.) $p_s(\vect{x})$, under 
rather general conditions the following relation holds (see 
Appendix~\ref{app:fdr}):
\begin{equation}
\label{eq:response}
 R^{kj}_t \equiv \lim_{\delta x^{(j)}_0 \to 0}\, \dfrac{\overline{\delta x^{(k)}_t}}{\delta x^{(j)}_0} =
-\Big\langle x^{(k)}_t \dfrac{\partial\ln p_s(\vect{x})}{\partial x^{(j)}}
\Big|_{\vect{x}_0} \Big\rangle\,,
\end{equation}
where the average $\langle \cdot \rangle$ is computed on the two-times joint 
p.d.f. ${p}^{(2)}_s(\vect{x}_t,\vect{x}_0)$. $R_t$ is the matrix of the linear 
response functions (at time $t$) of the considered system.

Eq.~\eqref{eq:response} shows the existence of a rigorous link among responses 
and correlations, provided that either the functional form of $p_s({\vect x})$ is 
known, or it can be inferred from data. Of course, in general the latter
will be a rather non-trivial task, at least in high-dimensional systems.

 \section{Linear Markov systems}
 \label{sec:linearmarkov}

In this Section we will limit ourselves to the study of linear stochastic 
processes of the form
\begin{equation}
\label{eq:evo}
 \vect{x}_{t+1}= A\vect{x}_t+ B\vect{\eta}_t
\end{equation}
where $A$ and $B$ are constant $n\times n$ matrices and the components of ${\bf 
\eta}_t$ are independent and identically distributed random variables with zero 
mean and unitary variances. The spectral radius of $A$ needs to be less than 1, 
in order for the dynamics not to diverge with time. In this case one has  
$\langle x^{(i)}\rangle=0\quad \forall i $. The following relation between the 
response matrix and the covariance matrix with entries $C_t^{kj}=\langle 
x^{(k)}_t x^{(j)}_0\rangle$ holds:
\begin{equation}
\label{eq:respcorr}
 R_t=C_t C^{-1}_0
\end{equation}
where $C^{-1}_0$ is the inverse of $C_0$~\cite{marconi08} (see 
Appendix~\ref{app:linear} for details). This result can be shown to hold also in 
cases with continuous time.

Following the idea of Green-Kubo formula, which allows to understand the average 
effect of an electric field on the current in terms of 
correlations~\cite{kubo_response,LiviPoliti}, a cumulative ``degree of 
causation'' $ x^{(j)} \to x^{(k)} $ can be introduced:
\begin{equation}
\label{eq:greenkubo}
 \mathcal{D}_{j \to k}= \sum_{t=1}^\infty R_t^{kj}  \, .
\end{equation}
 This quantity  characterizes the cumulative effect of the perturbation $\delta 
x_0^{k}$ on the variable $x^{(k)}$. In linear systems with discrete time, from 
the relation $R_t=A^t$ (see Appendix~\ref{app:linear}) it follows that
 \begin{equation}
 \label{eq:cuma}
  \mathcal{D}_{j \to k}=\left[A(I_n-A)^{-1}\right]^{kj}\,,
 \end{equation}
 $I_n$ being the $n \times n$ identity matrix. 
 Let us stress that a vanishing value of $\mathcal{D}_{j \to k}$ does not 
exclude causation between $x^{(j)}$ and $x^{(k)}$; indeed, since ${R}_t^{kj}$ 
can assume both positive and negative values, contributions with opposite signs 
in the sum~\eqref{eq:greenkubo} might eventually compensate and give a null result 
even in presence of a causal link.

\subsection{Interventional and observational causation}
Let us briefly discuss an important difference between the FR formalism and the 
other traditional approaches to the study of causation.  The formalism of 
response, as well as Pearl’s probabilistic interventional approach~\cite{PearlBook,ay08}, focuses on 
the effect of an active perturbation of the considered system, which is a 
typical physical procedure in experimental practice. In contrast, GC and TE 
pertain mainly to the observational approach, as they are related to 
the information exchange between degrees of freedom.
As mentioned in the Introduction, the intrinsic statistical fluctuations of the 
observed variables are not crucial to establish their cause-effect relation from 
a physical, interventional perspective, because they are not related to the active 
perturbation of the system and its effects. On the other hand, such fluctuations 
play a relevant role in the information-based, observational approach, since 
they concur to determine the statistics of the observed quantities, and this is 
relevant to our ability to make prediction.

To show the above point, let us consider model~\eqref{eq:evo} with
\begin{equation}
\label{eq:simplemodel}
 A=\dfrac{\sqrt{2}}{2}\begin{pmatrix}
1 & 1\\
0 & 1
\end{pmatrix}
\quad \quad
 B=\begin{pmatrix}
\sqrt{D_1} & 0\\
0 & \sqrt{D_2}
\end{pmatrix}\,.
\end{equation} 
The response function $R^{12}_{t=1}=A^{12}$ is equal to $\sqrt{2}/2$ and 
is independent of $D_1$ and $D_2$, as it is expected. Indeed, the amplitudes of 
the noise terms should not play any role in the cause-effect relations, from a 
physical perspective.

For direct comparison, let us compute now the GC and the TE for the same model. 
Suppose it generates a long time-series $\{x^{(1)}_t,x^{(2)}_t\}$: 
the evaluation of the observational casual link between 
$x^{(2)}$ and $x^{(1)}$ with the GC test requires to find
the best approximation of $\{x^{(1)}_t\}$ by the two alternative models
\begin{equation}
\label{eq:modelgp1}
 x^{(1)}_{t+1}=\alpha_1 x^{(1)}_t + \sqrt{\Delta_1} \xi_t
\end{equation} 
and
\begin{equation}
\label{eq:modelgp2}
 x^{(1)}_{t+1}=\alpha_2 x^{(1)}_t + \beta_2 x^{(2)}_t + \sqrt{\Delta_2} \xi_t\,,
\end{equation} 
where the coefficients $(\alpha_1,\Delta_1)$ and $(\alpha_2,\beta_2,\Delta_2)$ need to be optimally adjusted.
Once they are known, the quantity
\begin{equation}
 GC_{2 \to 1}=\ln\left(\dfrac{\Delta_1}{\Delta_2}\right)
\end{equation}
provides a measure of the increment in the predictability of $x^{(1)}$ when also 
the trajectory of $x^{(2)}$ is taken into account. 
In Appendix~\ref{app:gp}, we  
compute $\Delta_1$ and  $\Delta_2$ explicitly for 
model~\eqref{eq:simplemodel}, finding the final result
\begin{equation}
\label{eq:gpresult}
GC_{2 \to 1}=\ln \dfrac{1+4r+2r^2}{1+3r}\,,
\end{equation} 
where $r=D_2/D_1$.
Likewise, we can derive analytically the TE for model~\eqref{eq:simplemodel}. In this case, we need to evaluate the following expression:
\begin{equation}
 TE_{2 \to 1}=\left \langle \ln  \dfrac{p(x^{(1)}_{t+1}|x^{(1)}_{t},x^{(2)}_{t})}{p(x^{(1)}_{t+1}|x^{(1)}_{t})}  \right \rangle\,,
\end{equation} 
where the average is taken over the joint distribution $p(x_{t+1},x_t,y_t)$.
In Appendix~\ref{app:te} we show that
\begin{equation}
\label{eq:teresult}
TE_{2 \to 1}=\dfrac{1}{2} \ln \dfrac{1+4r+2r^2}{1+3r}\,.
\end{equation} 
The coincidence of TE and GC expressions, but for 
a factor $1/2$, is not incidental: indeed, the equivalence of the two 
quantities for linear regressive systems has been proved in 
Ref.~\cite{barnett09}. Both TE and GC depend on the ratio $r=D_2/D_1$ of the 
noise amplitudes: as mentioned at the beginning of this Section, this is 
consistent with the fact that they are related to predictability rather than to 
mechanistic causality, in contrast with response.

Let us stress that also in the response-theory approach one may define a 
observational-like causation estimator by rescaling correlations and responses 
with the standard deviations of the corresponding variables:
\begin{equation}
 \label{eq:rescaled}
 \tilde{C}_t^{kj}=\dfrac{1}{\sigma_k \sigma_j}C_t^{kj} \quad    \tilde{R}_t^{kj}=\dfrac{\sigma_j}{\sigma_k } R_t^{kj}\,.
 \end{equation}
 Since the quantities $\tilde{R}_t^{kj}$ are dimensionless, they can be used to 
compare the effect of different ``causes'' on a given variable. In the above 
discussed example, the rescaled response reads:
\begin{equation}
 \tilde{R}^{12}=\sqrt{\dfrac{1}{2}\left(3+\dfrac{1}{r}\right)^{-1}}\,.
\end{equation} 

\subsection{Linear response and correlations}
\begin{figure}
 \centering
 \includegraphics[width=.9\linewidth]{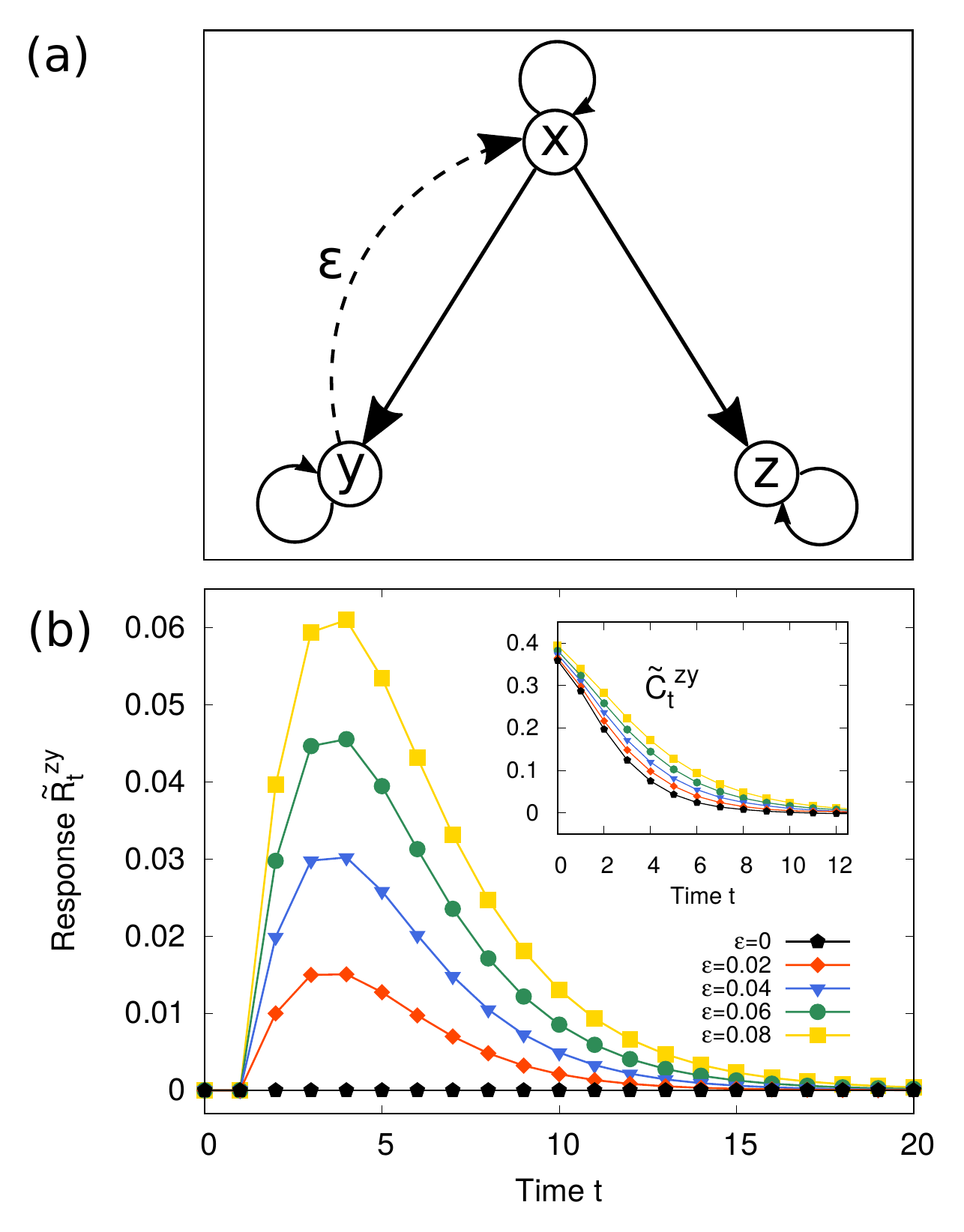}
 \caption{Spurious correlations and response. Panel (a) schematically 
represents the coupling scheme of Eq.~\eqref{eq:chocolate}, 
where solid arrows account for linear dependences with coefficient $a$, 
while the dashed arrow indicates the linear term multiplied 
by $\varepsilon$. 
Panel (b) shows the rescaled response (Eq.~\eqref{eq:rescaled}) of $z$ 
when $y$ is perturbed. 
The inset plots the corresponding correlations. 
Several values of $\varepsilon$ are considered; in all cases, 
$a=0.5$, $b=1$. 
Each plot has been obtained with an average over $10^5$ trajectories; 
responses have been computed inducing an initial perturbation 
$\delta y_0=0.01$.}
 \label{fig:1}
\end{figure}

To better understand the role of response in determining non-trivial causal 
links, let us examine a typical toy model in which the analysis of correlations
may lead to wrong conclusions.
We consider a 3-dimensional vector $\mathbf{x}=(x,y,z)$, whose evolution is 
ruled by a Gaussian, linear stochastic dynamics at discrete times:
\begin{subequations}
\begin{align}
  x_{t+1}=&a x_t + \varepsilon y_t  + b \eta^{(x)}_t \\
y_{t+1}=&a x_t + a y_t + b \eta^{(y)}_t \\
z_{t+1}=&a x_t + a z_t + b \eta^{(z)}_t
\end{align}
\label{eq:chocolate}
\end{subequations}
where  $\eta^{(x)}, \eta^{(y)}, \eta^{(z)}$ are independent Gaussian processes 
with zero mean and unitary variance, while $a$, $\varepsilon$ and $b$ are 
constant parameters. 
The situation is graphically represented in 
Fig.~\ref{fig:1}(a). The case $\varepsilon=0$ is a minimal example in which the 
behavior of two quantities, $y$ and $z$, is influenced by a common-causal 
variable $x$; as a consequence, $y$ and $z$ are correlated even though they are 
not in causal relationship (black graph in the inset of 
Fig.~\ref{fig:1}(b)). 
The same mechanism may be identified in many situations 
in which surprising functional dependences arise, as that between 
the number of Nobel laureates of a country and its chocolate consumption per 
year~\cite{messerli12}: in this specific case, both quantities may be expected 
to be influenced by the gross domestic product of the nation.

According to our definition, in order to decide whether there is a causal 
relation between $y$ and $z$, one has to perturb $y$ at time $0$ and measure the 
average variation $\delta z_t$ for $t>0$. 
Let us briefly comment on the optimal choice for the intensity 
of the perturbation. As a general rule, 
$\delta y$ should be small with respect to the typical values of the variable 
$y$, since the linear response theory requires an expansion for small 
values of $\delta y$ (see Appendix~\ref{app:fdr});
on the other hand, if $\delta y$ is too small, a large number of experiments 
will be needed to get reliable averages over the stochastic realizations 
of the noise. 
Here and in the following examples, we took $\delta y \simeq O(10^{-2})$;  
however, since the dynamics of this example is linear, the 
results of Appendix~\ref{app:fdr} are exact and there is actually no need to 
choose $\delta y$ small.

The result for $\varepsilon=0$ is shown in 
Fig.~\ref{fig:1}(b), black curve: not surprisingly, $\tilde{R}_t^{zy}=0$ for all 
$t>0$. The situation completely changes if we introduce a small feedback 
$\varepsilon\ne 0$ from $y$ to $x$, which will eventually result in a causal 
link between $y$ and $z$. As Fig.~\ref{fig:1}(b) shows, the corresponding 
response function correctly reveals that the behavior of $z$ starts to be 
influenced by a perturbation of $y$ after $t=2$ time steps, and that the 
intensity of such causal influence roughly scales with $\varepsilon$. 

None of these conclusions could have been drawn from the mere analysis of the 
correlation functions, reported in the inset of Fig.~\ref{fig:1}(b). However, 
for linear Markov systems, formula~\eqref{eq:respcorr} allows 
the response function to be found by simple operations on the covariance 
matrix, i.e. by a suitable manipulation of time correlations.

It can be shown~\cite{barnett09,palus07, liang16} that in linear systems 
also GC, TE and related quantities can be eventually
reduced to functions of correlations, but in general their derivation 
may be much more involved than that based on response theory. 
In studying the above example, an important \textit{caveat} has to be 
bore in mind: when dealing with more than two variables, 
in order to get insightful results, we need to use 
\textit{conditional} GC and TE~\cite{barrett10}. 
This fact can be understood by looking at the causal link between $y$ and $z$ 
with a time-lag of 1 step, which is expected to be null from a physical 
perspective, since no action on $y_t$ will have consequences on 
$z_{t+1}$ in our model. 
The ``naive'' TE
\begin{equation}
 TE_{y\to z}=\left\langle \ln \dfrac{p(z_{t+1}|z_{t}, y_{t})}{p(z_{t+1}|z_{t})} \right\rangle
\end{equation} 
will be in general different from zero, because the knowledge of $y_t$ provides 
indirect information about $x_t$ (the two variables are not independent), and 
the possibility to forecast the value of $z_{t+1}$ is improved. 
The problem is solved by considering the conditional TE
\begin{equation}
\label{eq:TEcond}
 TE_{y\to z|x}=\left\langle \ln \dfrac{p(z_{t+1}|z_{t},x_t, y_{t})}{p(z_{t+1}|z_{t}, x_t)} \right\rangle\,;
\end{equation}
in this case, the conditional probabilities at the numerator and 
denominator are equal, in fact the knowledge of $y_t$ does not provide 
additional information about $x_t$, which is already known. 
Similar considerations hold for the GC analysis. 
However, let us stress that the FR formalism provides a handy 
method to deal with many variables at the same time, as in the linear cases 
the problem reduces to the computation of 1-step correlations and matrix 
operations. 
The TE approach, instead, requires the evaluation from data of conditioned 
probabilities as those appearing in Eq.~\eqref{eq:TEcond}, which may 
be a non-trivial task as soon as the number of conditioning variables 
is larger than 1 or 2.

\begin{figure}
\centering
\includegraphics[width=.9\linewidth]{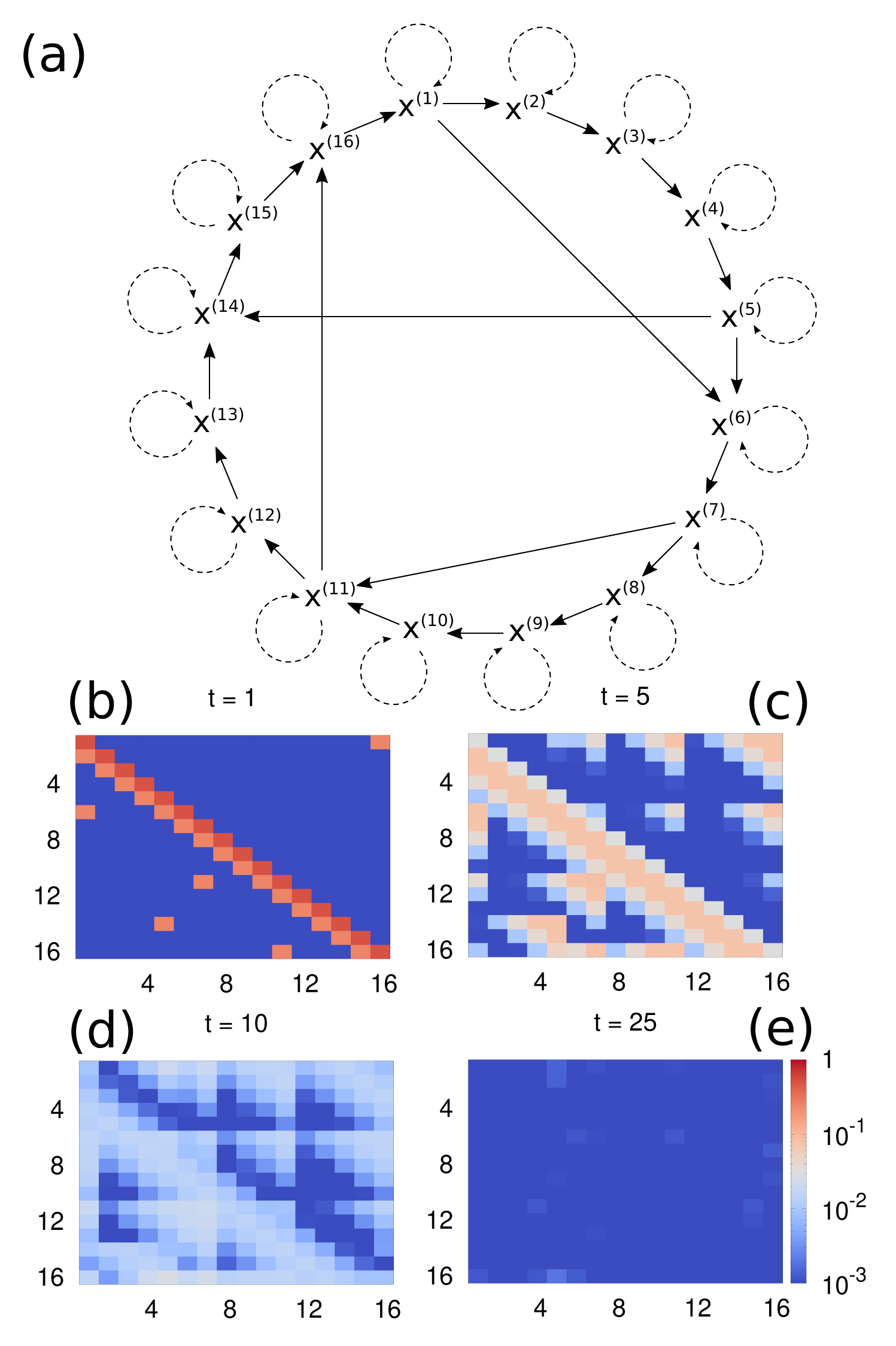}
\caption{Response in multidimensional linear systems. 
Panel (a) schematically shows an example of interactions scheme for
a model of the form~\eqref{eq:evo}. 
Each solid arrow represents a linear interaction coefficient $0.25$, 
each dashed arrow stands for an auto-interaction term $0.5$.
 The response matrix $R_t$ is represented in panels (b)-(e) for
 different values of $t$, according to the color scheme in panel (e). 
Here $b=1$, and correlations have been obtained 
by averaging over $10^6$ trajectories.}
 \label{fig:2}
\end{figure}

\begin{figure}
 \centering
 \includegraphics[width=.9\linewidth]{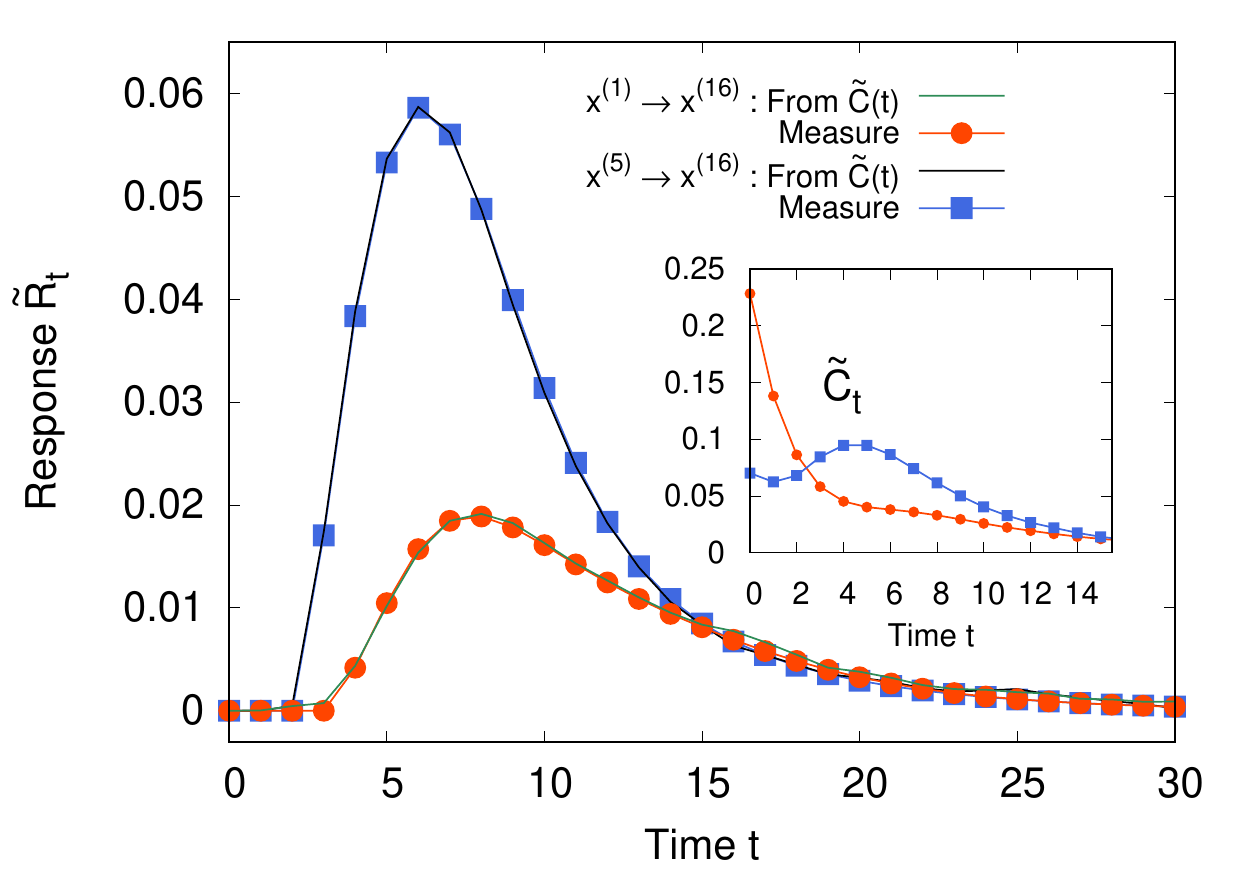}
 \caption{Causation and correlation in multidimensional linear systems. Response 
functions (main plot) and correlations (inset) for the causal links $x^{(1)}\to 
x^{(16)}$ and $x^{(5)}\to x^{(16)}$ of the model described by  
Fig.~\ref{fig:2}.  Simulated responses (perturbation amplitude $\delta x_0=0.01$) 
are compared with formula~\eqref{eq:respcorr}.}
\label{fig:3}
\end{figure}

\subsection{``Direct'' causation and modeling via response theory}

A typical problem in the study of a complex system is that of inferring 
the strength of its  links, assuming that the dynamics is of the form~\eqref{eq:evo}; 
in other terms, one can be interested in inferring the matrix $A$ from the 
analysis of long time series $\{ x_t ^{(i)} \}$,  $i=1,\ldots,n$, 
$t=1,2,\ldots,T \gg 1$. A situation of this kind is usually faced, e.g., in the 
study of complex proteins~\cite{Kaneko,Piazza}. In these cases one is mostly 
interested in the ``direct'' causation links between the variables, which allow 
to understand the structure of the system and the matrix $A$~\cite{sun15}; this 
can be done again by mean of response theory, which relates the response 
function to the propagator of the dynamics. In particular, by recalling that 
$R_t$ and $A$ are simply related by $R_t=A^t$, one has that $A=R_1$. 
An example is shown in Fig.~\ref{fig:2}; the matrix $A$ which rules the 
dynamics is graphically represented by panel (a). 
In panels (b)-(e) the matrix $R_t$ is shown 
as reconstructed from time correlations, for different values of $t$. 
As expected, for $t=1$ the response matrix equals $A$, and it is possible to 
infer all (oriented) causal links. 
For $t>1$, $R_t^{kj}$ provides information on the indirect influence of 
$x^{(j)}$ on $x^{(k)}$, i.e. including effects which would not have been 
present in a system composed by $x^{(j)}$ and $x^{(k)}$ only.

However, the response formalism is able to give, with minimal effort, much more 
information on the studied system. 
In particular, it is especially suitable to determine in a rather simple 
way also ``indirect'' causation. 
It is quite natural to say that there exists an indirect causation 
relationship $ x^{(j)} \to x^{(k)} $ if there exist an oriented path on the graph connecting $j$ with 
$k$, i.e. there is (at least) a sequence of length $m-1$ $(i_1, i_2,\ldots, 
i_{m-1})$ such that
\begin{equation}
A_{i_1,j} \neq 0 \,,\; A_{i_2,i_1} \neq 0 \,,\;\ldots\,,\; A_{k,i_{m-1}} \neq 0\,.
\end{equation} 
From the time series $\{x_t ^{(i)} \}$,  $i=1, ... ,n$, we can compute the 
correlation functions and, using Eq.~\eqref{eq:respcorr}, the response matrix. 
The entries $R_t^{kj}$ allow the understanding of the structure of the graph 
(i.e. the matrix $A$) and the causation relationship $ x^{(j)} \to x^{(k)} $. If 
$R_t^{kj}=0$ for any $t>0$,  the causation link is missing, whereas
if $R_t^{kj}=0$  for $t\le m-1$ and $R_t^{kj}\neq 0$ for $t\ge m$, this means that there exist at 
least a path of length $m$ connecting $j$ with $k$. 
Fig.~\ref{fig:3} reports
two examples of response functions ($\tilde{R}_t^{16,1}$ and 
$\tilde{R}_t^{16,5}$) for the model described in Fig.~\ref{fig:2}. 
It can be verified that, in both cases, the first non-zero value of the 
responses obtained after a number of time-step equals the length of the 
minimum oriented path connecting the considered variables. 
Again, the relative effect of the 
variables $x^{(1)}$ and $x^{(5)}$ on $x^{(16)}$ could not have been simply 
deduced from the correlation functions, reported in the inset of 
Fig.~\ref{fig:3}.

Let us just mention that the same reasoning can be easily extended to 
stochastic processes with continuous time of the form
\begin{equation}
 \dot{\vect{x}}=-F\vect{x}+ B \vect{\xi}\,,
\end{equation} 
where $F$ and $B$ are $n\times n$ matrices. 
The eigenvalues of $F$ have positive real part and $\vect{\xi}$ is a 
$n$-dimensional, delta-correlated normalized Gaussian noise. 
In this case it can be shown~\cite{marconi08} that $R_t=\exp(-Ft)$, so 
that inferring $F$ from the study of the response functions is again possible, 
either by considering the matrix $I_n-R_t$ for $t\to 0$, where 
$I_n$ denotes the $n \times n$ identity matrix, or by the continuous-time 
version of Eq.~\eqref{eq:cuma},
\begin{equation}
\mathcal{D}_{j\to k}=[F^{-1}]^{kj}\,.
\end{equation}

\begin{figure}
 \centering
 \includegraphics[width=.9\linewidth]{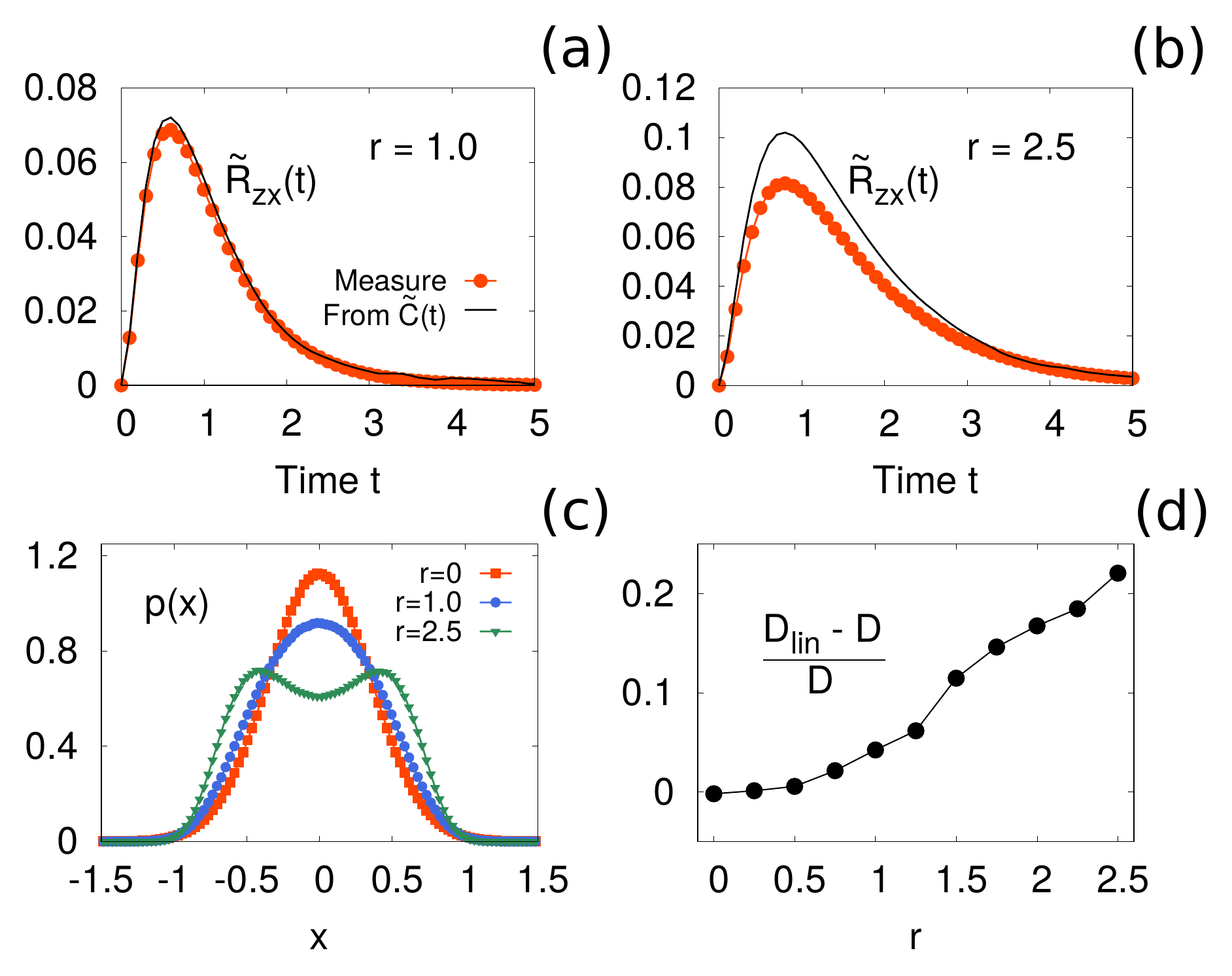}
 \caption{Response function in the nonlinear dynamics~\eqref{eq:nonlin}. Panels 
(a) and (b) show $\tilde{R}_t^{zx}$ for two different values of the nonlinearity 
$r$, compared with formula~\eqref{eq:respcorr} (which is valid in the linear 
case). Panel (c) shows the p.d.f. of the variable $x$, for different values of 
$r$. In panel (d) the relative error is shown between the integrated response 
$\mathcal{D}_{x \to z}$, computed from the measured $\tilde{R}_t^{zx}$, and its 
approximation $\mathcal{D}_{lin}$, where the proxy responses of 
formula~\eqref{eq:respcorr} are considered. For the numerical simulations a 
stochastic Heun integrator has been used (see e.g. Ref.~\cite{rumelin82}), 
choosing $b=1$, $k=1$, and a time-step $\Delta t=0.001$; perturbation amplitude 
$\delta x=0.01$; each plot has been obtained by averaging over $10^6$ 
trajectories.}
\label{fig:4}
\end{figure}

\section{Tackling the general problem
\label{sec:generalization}}

In this Section we discuss the difficulties encountered when 
trying to infer causal relations in more general situations, 
as in non-linear systems and in cases where not all relevant 
variables are accessible. 
While, in the former case, the FD theory is still applicable in principle, 
and linear approximations provide quite good results, in the latter the 
lack of information is a major obstacle to the understanding of the causal 
links.

\subsection{Non-linear systems}

As an example of non-linear dynamics, let us now consider a system 
composed by three interacting particles in one dimension moving 
under the action of an external non-harmonic potential. 
We assume an overdamped dynamics, so that 
the state of the system $\vect{x}=(x,y,z)$ evolves as   
\begin{subequations}
\label{eq:nonlin}
\begin{align}
 \dot{x}&=-U'(x)-k(x-y)+b \xi^{(x)}\\
 \dot{y}&=-U'(y)-k(y-x)-k(y-z)+b \xi^{(y)}\\
 \dot{z}&=-U'(z)-k(z-y)+b \xi^{(z)}
 \end{align}
\end{subequations}
with
\begin{equation}
 U(x)=(1-r)x^2+rx^4\,,
\end{equation}
where $k$ and $b$ are constants, $\vect{\xi}$ is a delta-correlated Gaussian 
noise and $r$ is a parameter which determines the degree of nonlinearity of the 
dynamics: when $r=0$ the external potential $U$ is harmonic, 
while for $r>1$, it takes a double-well shape. 
We are interested in studying how accurate Eq.~\eqref{eq:respcorr} is in 
predicting the response function. 
Eq.~\eqref{eq:response} implies that the general treatment of cases in which 
the invariant p.d.f. is not Gaussian would require 
(i) a careful estimation of the functional form of the joint p.d.f. of 
\textit{all} variables of the system and 
(ii) the knowledge of all correlation functions resulting from the r.h.s. of 
Eq.~\eqref{eq:response}. 
However Fig.~\ref{fig:4} shows that if the nonlinear 
contribution to the dynamics is small enough, the ``linearized'' 
response~\eqref{eq:respcorr} still gives a meaningful information about the 
causal relations between the variables of the system. 
In particular, Fig.~\ref{fig:4}(d) reports the relative error that one 
makes by computing $\mathcal{D}_{x \to y}$, defined by 
Eq.~\eqref{eq:greenkubo}, with the linear approximation 
Eq.~\eqref{eq:respcorr}. 
We observe that the error is rather bounded even for $r \simeq O(1)$, 
when the joint p.d.f. is quite far from a multivariate Gaussian. 
This fact has a quite clear mathematical interpretation. 
To show that, we consider a system described by the time-dependent vector 
$\vect{x}$, ruled by some unknown stochastic dynamics. 
The system is initially in the state $\vect{x}_0$, and the dynamics will 
evolve it to some other state $\vect{x}_t$ after a time interval $t$, 
where $\vect{x}_t$ will, in general, depend both on the initial condition 
and on the particular realization of the stochastic noise. 
If we repeat this kind of observation many times along a 
trajectory, assuming that the dynamics of the considered system is ergodic, we 
can collect many pairs $(\vect{x}_0,\vect{x}_t)_i$. The best linear 
approximation to predict $\vect{x}_t$ from $\vect{x}_0$ will be of the form
\begin{equation}
\label{eq:approxlin}
 \vect{x}_t \simeq L_t \vect{x}_0 + \vect{\zeta}_t
\end{equation} 
where $\vect{\zeta}_t$ is a vector of random variables with zero mean, 
independent of $\vect{x}_0$. The structure of Eq.~\eqref{eq:approxlin} is the 
same as that of Eq.~\eqref{eq:xt}. Reasoning as in Appendix~\ref{app:linear}, 
one finds the linear regression formula (see also Ref.~\cite{barnett09})
\begin{equation}
 L_t \simeq C_t C_0^{-1}\,.
\end{equation}  
As a consequence, the $R_t$ matrix that one might compute in nonlinear systems 
by using the ``wrong'' relation $R_t=C_tC_0^{-1}$ is actually the response 
associated to the process~\eqref{eq:approxlin}, which is the best linear 
approximation of the considered transformation $\vect{x}_0 \to \vect{x}_t$. Let 
us notice that for this result to hold, we do not have to assume any particular 
dependence of $L_t$ on time.

\subsection{Systems with hidden variables: failure of ``embedding'' strategies}

Let us conclude by discussing the rather common situation in which we do not 
know the whole state vector ${\mathbf x}$ of the system, 
but we only have access to the time series of two variables, 
$\{x_t^{(j)} \}$ and $\{ x_t^{(k)} \}$. 
In order to show the basic problems in inferring causation, 
let us refer again to the system~\eqref{eq:chocolate}, assuming that 
only the times series of $y$ and $z$ are available.

\begin{figure}
 \centering
 \includegraphics[width=.9\linewidth]{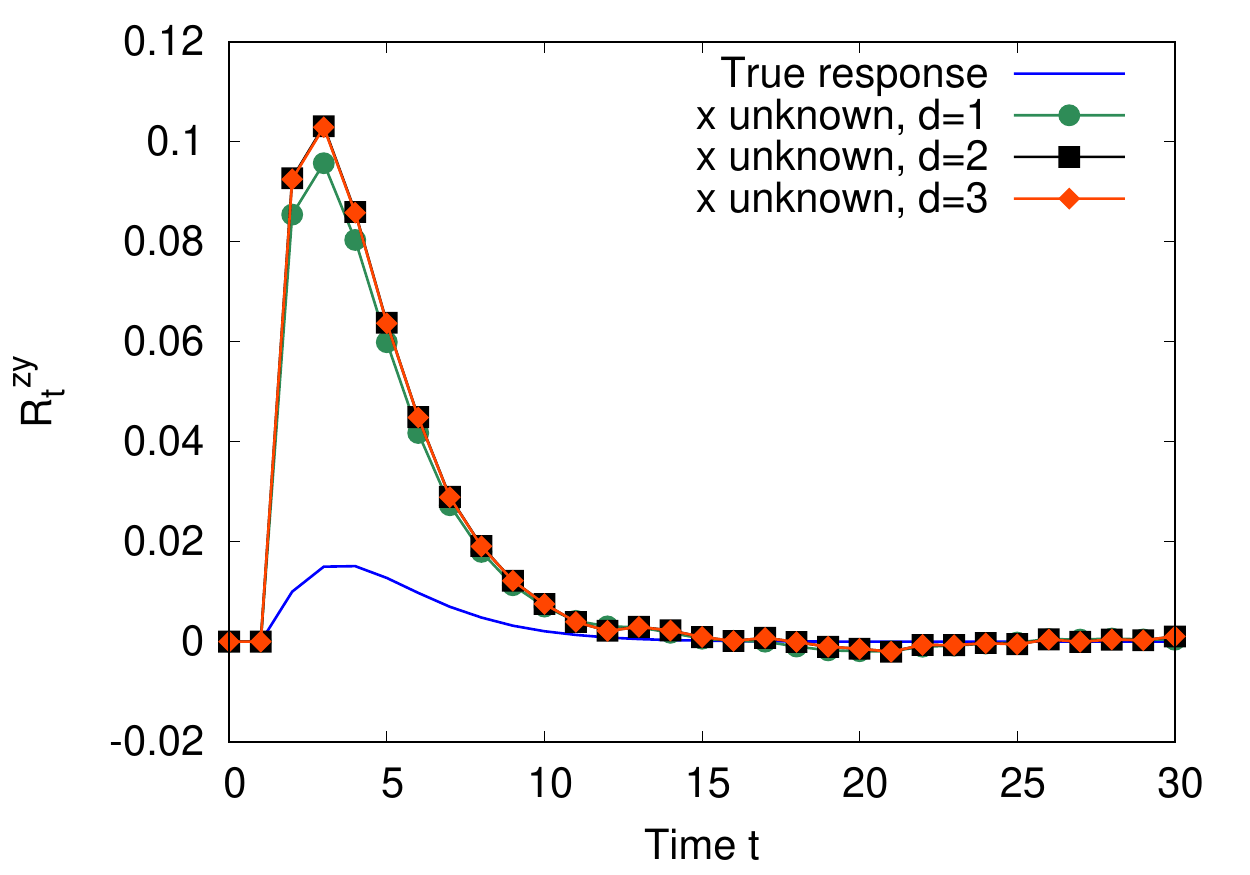}
 \caption{Estimating the response function of model~\eqref{eq:chocolate}, with 
$a=0.5$, $\varepsilon=0.02$, $b=1$. The blue solid line shows the actual 
response function $R^{zy}_t$, measured from simulations; the other curves, 
marked with different symbols, represent the results obtained from the only 
knowledge of correlations between $y$ and $z$, as if $x$ was not part of the 
system, with different embedding dimensions. All curves have been obtained with 
an average over $10^5$ trajectories.}
\label{fig:5}
\end{figure}

The first attempt to detect the $y-z$ causation can be to consider the reduced 
vector  $\vect{\Gamma}_t^{(2)} = (z_t,y_t)$, assuming that it properly describes 
the system, and to use formula~\eqref{eq:respcorr} in this 2-dimensional 
space. 
This simple approach leads to wrong results: as shown in Fig.~\ref{fig:5} 
(green circles), the computed ``response'' function is completely different 
from the real $R^{zy}_t$ (blue solid line). 
This is not surprising at all, since  $\vect{\Gamma}_t^{(2)}$ 
does not contain enough information about the state of the system, 
and therefore the dynamics is not Markov. 
A tempting strategy, inspired by Takens' ``embedding'' approach in the context 
of deterministic dynamical systems~\cite{Takens81,kantz03}, suggests to try 
a reconstruction of a vector which completely describes the state of the 
system, by exploiting the knowledge of past values of $y_t$ and $z_t$. 
Basically, the idea is to introduce the vector
\begin{equation}
\vect{\Gamma}_t^{(2d)}=(y_t,y_{t-1},..., y_{t-d+1}, z_t,z_{t-1},..., z_{t-d+1}) \,,
\end{equation}
and repeat the analysis on this $2d$-dimensional system for increasing values 
of $d$. 
For deterministic dynamical systems, if $d$ is large enough, 
the vector $\vect{\Gamma}_t^{(2d)}$ can be proved to have an autonomous 
dynamics, so we might expect that in the context of stochastic processes 
it would follow a Markov evolution rule. 
If this were the case, we could apply formula~\eqref{eq:respcorr} 
to $\vect{\Gamma}_t^{(2d)}$ and infer all the causal links. 
Unfortunately, Fig.~\ref{fig:5} shows in a rather convincing way that 
increasing the embedding dimension $d$ does not lead to any improvement: on 
the contrary, choosing $d>1$ can even determine, as in the considered example, 
a worse estimation of the response function. Similar results would have been 
observed with different choices of the embedding protocol.

The embedding fails for generic random process because, at variance with 
deterministic cases, the knowledge of previous values of certain observables is not equivalent, 
in general, to the knowledge of the entire state of the system. 
To clarify this point, let us consider a dynamical system $\vect{x}_t$ composed of $n$ variables 
$(x^{(1)},x^{(2)},...,x^{(n)})$, ruled by some autonomous dynamics in discrete 
time:
\begin{equation}
\label{eq:evo_det}
\vect{x}_{t+1}=\vect{f}(\vect{x}_{t})\,,
\end{equation} 
where $\vect{f} : \mathbb{R}^n \mapsto \mathbb{R}^n$, and we know that there 
exists a unique solution at any time. It is quite obvious that the 
$n$-dimensional vector obtained with the embedding protocol:
\begin{equation}
\mathbf{\Gamma}^{(n)}_t=(x^{(1)}_{t}, x^{(1)}_{t-1}, ..., x^{(1)}_{t-n+1})
\end{equation} 
gives as much information as the vector $\vect{x}_t$, see e.g. Ref~\cite{Takens81,kantz03}.

Let us now consider a non autonomous version of~\eqref{eq:evo_det},
\begin{equation}
\vect{x}_{t+1}=\vect{f}(\vect{x}_{t})+\vect{g}(t)\,, 
\end{equation} 
where  $\vect{g}:\mathbb{R} \mapsto \mathbb{R}^n$ is the vector of $n$ periodic 
functions with period $T$. The system can be mapped into an autonomous system by 
introducing a new variable, say $w$, such that
\begin{equation}
\label{eq:w}
\begin{cases}
 w_0=0\\
 w_{t+1}=w_t+1-T\lfloor (w_t+1)/T\rfloor\,,
\end{cases}
\end{equation}
where $\lfloor y \rfloor$ stands for the integer part of $y$. With this 
definition, $w_t\in[0,T)$ and $\vect{g}(t)=\vect{g}(w_{t})$, because of its 
periodicity. Similarly, if $\vect{g}(t)$ is the linear combination of periodic 
functions with $k$ (incommensurable) periods $T_1,...,T_k$, 
system~\eqref{eq:evo_det} can be mapped into an autonomous system by introducing 
$k$ variables $w^{(1)},...,w^{(k)}$ of the form~\eqref{eq:w}.

Since a random term can be seen as the superposition of an infinite number of 
periodic functions with incommensurable frequencies, it is straightforward to 
understand that in a generic system perturbed by a random forcing, for any 
finite $d$,  the vector $\vect{\Gamma}^{(d)}_s$ cannot be able to describe 
completely the state of the original system. 
In particular, no reliable information about the response function of 
the original system can be deduced by applying the fluctuation-response 
relation to it.

This implies that to infer causation from time correlations in 
a stochastic dynamics, 
we actually need to know the trajectories of all the variables 
which are relevant to the dynamics of $y$ and $z$.

\section{Conclusions}
\label{sec:conclusions}

Using some tools from the FR theory of out-of-equilibrium statistical 
mechanics, we have introduced a way of characterizing causation 
between two variables, whose physical interpretation is rather 
straightforward. 
The basic idea  of this 
proposal is that $x^{(j)}$ has a causal effect on $x^{(k)}$ after a time 
interval $\Delta t>0$ if a perturbation of $x^{(j)}$ at time $t$ induces some 
change on $x^{(k)}$ at time $t+\Delta t$. 
In this sense, our definition is reminiscent of the interventional 
framework developed by  Pearl, in which 
causation is detected by observing the effects of an action on the system. 
Other approaches to detect causation, as those related to GC and TE, are 
based on the idea that causation is associated to information, 
i.e. $x^{(j)}$ has an effect 
on $x^{(k)}$ if the  knowledge of $x^{(j)}$ helps the prediction of $x^{(k)}$. 
At a first glance, the choice between observational 
or interventional approaches may seem only a matter of taste; instead the two 
methods present important differences, both at qualitative and quantitative 
level.

Bearing in mind the above definition, we describe a practical method to 
understand causal links between the variables of a system by looking at 
time-series of data. Despite the (correct) common-wisdom statement that 
correlation does not imply causation, we have shown  
that, at least in multi-dimensional linear 
Markov process, the presence/absence of causation between variables can 
be inferred by a proper employ of (all) time correlations. 
The FR formalism can be used to find ``direct'' causal links between 
variables at a given time, and therefore to build linear models based on 
these findings, as well as to introduce a ``degree of causation'' 
cumulative in time. 
The  physical interpretation of this indicator is quite natural and 
reminds the Green-Kubo formula for the electric (or thermal) conductivity.

From a computational point of view the practical implementation of our method is 
quite easy, much simpler than GC and TE, whose application becomes elaborate in 
high dimensional systems. In a generic nonlinear dynamics, even though an exact 
relation between response functions and certain correlators (whose specific 
shape depends on the invariant probability distribution) always exists, its 
explicit form may be very convoluted. However, we have shown that the protocol 
that holds for the linear case still represents a useful proxy also in presence 
of weak nonlinear terms.

Serious difficulties arise instead in the case of hidden variables, i.e. when 
the access to the vector ${\bf x}$ describing the state of 
the system is partial. The tempting idea to use an ``embedding'' methodology to 
reconstruct the proper complete phase space, at variance with 
deterministic systems, does not work, in general, for stochastic processes. 
Let us stress that this impossibility is not due to mere practical 
difficulties, as the limited length of the time series or the high 
dimension of the system. 
It seems to us that the only possible way to understand causation from 
data is to guess the proper set of variables which describe, 
at least within a certain accuracy, the complete system according to a 
Markov rule. 
The above limitation is always present in any purely inductive approach, 
i.e. in all cases where, without a fair theoretical framework, 
one tries to infer the essence of a system (or to build an effective model) 
just from  data. 
Caveats on this topic had been already expressed by Onsager and 
Machlup~\cite{onsager53}, and Ma~\cite{ma85}, in a rather vivid way; 
unfortunately, those wise warnings are often disregarded.

\begin{acknowledgements}

The Authors thankfully acknowledge useful discussions with Erik Aurell. 
This work is part of MIUR-PRIN2017 
\textit{Coarse-grained description for non-equilibrium 
systems and transport phenomena (CO-NEST)} whose partial financial support is 
acknowledged. 
\end{acknowledgements}

\appendix

\section{Linear response in a nutshell}
\label{app:fdr}

Just for the sake of self-consistency, here we recall the main ideas and results of linear response theory. A more detailed exposition can be found, for instance, in Ref.~\cite{marconi08}.
Consider a Markov process $\vect{x}_t=(x_t^{(1)}, ..., x_t^{(n)})$ whose invariant p.d.f. $p_s(\vect{x}_t)$
is smooth and nonvanishing. Given a (small) perturbation $\delta \vect{x}_0 = (\delta x_0^{(1)}, ..., \delta x_0^{(n)})$ at time $t=0$, we want to understand its effects at time $t$ by measuring the difference between
 the vector $\vect{x}_t$ in the original dynamics and in the perturbed one, 
on average. More precisely, we want to compute
 \begin{equation}
 \overline{ \delta x^{(k)}_t} = \langle x^{(k)}_t \rangle_p -\langle x^{(k)}_t \rangle\,,
 \end{equation} 
where $ \langle \cdot \rangle_p$ and $ \langle \cdot \rangle$ indicate the average over many realizations of the perturbed and of the original dynamics, 
respectively.

We can compute explicitly the average $\langle x^{(k)}_t \rangle_p$ by noticing that the joint p.d.f. in the perturbed case can be written as
\begin{equation}
p_{pert}(x^{(k)}_t,\vect{x}_0)=p_s(\vect{x}_0)p_s(x^{(k)}_t| \vect{x}_0+\delta \vect{x}_0)\,,
\end{equation} 
where the stationary conditional probability accounts for the effect of 
the perturbation at time $t=0$.
As a consequence, the above average can be written as:
\begin{equation}
 \begin{aligned}
   \langle x^{(k)}_t \rangle_p=&\int d \vect{x}_0 dx_t^{(k)}\,p_s(\vect{x}_0)p_s(x_t^{(k)}|\vect{x}_0+\delta \vect{x}_0)x_t^{(k)}\\
   =&\int d \vect{x}_0 dx_t^{(k)}\,p_s(\vect{x}_0-\delta \vect{x}_0)p_s(x_t^{(k)}|\vect{x}_0)x_t^{(k)}\\
   \simeq \langle x^{(k)}_t \rangle &- \sum_j \delta x^{(j)}_0 \int d \vect{x}_0 dx_t^{(k)}\,\dfrac{\partial p_s(\vect{x}_0)}{\partial x^{(j)}_0}p_s(x_t^{(k)}|\vect{x}_0)x_t^{(k)}\,,
 \end{aligned}
\end{equation} 
where in the second line we have made a shift of the integration 
variables: $\vect{x}_0 \to \vect{x}_0-\delta \vect{x}_0 $.
From the above equation one easily finds
\begin{equation}
\label{eq:appresponse}
 \overline{ \delta x^{(k)}_t} \simeq -\sum_j \left\langle x^{(k)}_t\dfrac{\partial \ln p_s(\vect{x})}{\partial x^{(j)}}\Big|_{\vect{x}_0}\right\rangle\delta x_0^{(j)}\,,
\end{equation} 
whence Eq.~\eqref{eq:response}. 
The above formula can be generalized to a generic observable $\mathcal{F}(\vect{x}_t)$ as
\begin{equation}
 \overline{\delta \mathcal{F}(\vect{x}_t)}\simeq -\sum_j \left\langle \mathcal{F}(\vect{x}_t) \dfrac{\partial \ln p_s(\vect{x}_t)}{\partial x^{(j)}} \Big |_{\vect{x}_0} \right\rangle \delta x_0^{(j)}\,.
\end{equation} 

Let us notice that Eq.~\eqref{eq:appresponse} is valid under rather general 
hypotheses; in particular, in its derivation no assumption of detailed 
balance is used, meaning that Eq.~\eqref{eq:appresponse} also holds 
for out-of-equilibrium systems in stationary states.

\section{Response in linear systems}
\label{app:linear}

According to the definition of causation we followed in the paper, the variable $x^{(j)}$ influences $x^{(k)}$ if and only if
some smooth function $\mathcal{F}(x)$ exists such that
\begin{equation}
 \label{eq:resp_obs_app}
 \dfrac{ \overline{\mathcal{F}(\tilde{x}^{(k)}_t)-\mathcal{F}(x^{(k)}_t)}}{\delta x^{(j)}_0} \ne 0 \quad \quad \quad \mbox{for some }t>0\,,
\end{equation}
where $\delta x^{(j)}_0=\tilde{x}^{(j)}_0-x^{(j)}_0$ is the perturbation operated on $x^{(j)}$ at time $t=0$, 
and $\tilde{\vect{x}}_t$ represents the perturbed dynamics.

In the following, we want to show that this causal relation between 
two variables can be understood by only considering $\mathcal{F}(x)=x$ 
as far as linear Markov systems are concerned.
In order to show that, let us first recall that linear response theory 
allows to rewrite the l.h.s. of Eq.~\eqref{eq:resp_obs_app} as~\cite{marconi08}
\begin{equation}
 \label{eq:formulang}
\dfrac{ \overline{\delta \mathcal{F}(x^{(k)}_t)}}{\delta x^{(j)}_0} = -\Big\langle \mathcal{F}(x^{(k)}_t)  \dfrac{\partial\ln p_s(\vect{x})}{\partial x^{(j)}}
\Big|_{\vect{x}_0} \Big\rangle\,,
\end{equation} 
assuming that the considered process admits a smooth invariant distribution $p_s(\vect{x})$.
Let us now consider a $n$-dimensional system $\vect{x}_t=(x^{(1)}_t,...,x^{(n)}_t)$, 
whose dynamics is ruled by a discrete-time, stochastic linear evolution
\begin{equation}
\label{eq:evo_app}
 \vect{x}_{t+1}=A \vect{x}_{t} + B \vect{\eta}_t
\end{equation}
where $A$ and $B$ are $n \times n$ matrices and $\vect{\eta}_t$ is a $t$-dependent 
vector of delta-correlated random variables with zero mean. 
Eq.~\eqref{eq:evo_app} can be iteratively solved, leading to
\begin{equation}
\label{eq:xt}
 \vect{x}_t=A^t\vect{x}_0+\sum_{s=0}^{t-1}A^{t-s-1}B\vect{\eta}_s\,;
\end{equation}
as an immediate consequence, a simple relation holds between 
correlations and matrix $A$, namely
\begin{equation}
\label{eq:corr}
 \langle \vect{x}_t \vect{x}^{T}_0\rangle = A^t \langle \vect{x}_0 \vect{x}^{T}_0\rangle\,.
\end{equation} 

On the other hand, for this kind of systems the r.h.s. 
of Eq.~\eqref{eq:formulang} reads
\begin{equation}
\begin{aligned}
  &-\int d \vect{x}_0 d x_t^{(k)} \mathcal{F}(x^{(k)}_t) \dfrac{\partial \ln p_s(\vect{x}_0)}{\partial x_0^{(j)}}p_s(x_t^{(k)}|\vect{x}_0)p_s(\vect{x}_0)\\
  &=-\int d \vect{x}_0 d x_t^{(k)} \mathcal{F}(x^{(k)}_t) \dfrac{\partial p_s(\vect{x}_0)}{\partial x_0^{(j)}}p_s(x_t^{(k)}|\vect{x}_0)\\
 &=-\int d \vect{x}_0 d x_t^{(k)} \mathcal{F}(x^{(k)}_t)  p_s(\vect{x}_0) [A^t]^{kj}\dfrac{\partial p_s(x_t^{(k)}|\vect{x}_0)}{\partial x_t^{(k)}}
\end{aligned}
\end{equation}
where we have indicated by $p(x_t^{(k)}|\vect{x}_0)$ the probability density of $x_t^{(k)}$ conditioned to the initial state of the system
$\vect{x}_0$. The second equality is obtained with an integration by parts with respect to the variable $x_0^{(j)}$, 
bearing in mind that the last derivative can be switched from $x_0^{(j)}$ to $x_t^{(k)}$, because $p(x_t^{(k)}|\vect{x}_0)$ depends on $x_t^{(k)}$ and $x_0^{(j)}$ only through the linear combination $x_t^{(k)}-\sum_i[A^t]^{ki}x_0^{(i)}$.
Integrating again by parts, this time with respect to  $x_t^{(k)}$, one finally obtains
\begin{equation}
\label{eq:fgen}
 \dfrac{ \overline{\delta \mathcal{F}(x^{(k)}_t)}}{\delta x^{(j)}_0} =\langle \mathcal{F}\,' \rangle [A^t]^{kj} \,.
\end{equation} 
Calling $R_t$ the matrix of linear responses with the choice $\mathcal{F}(x)=x$, and taking into account Eq.~\eqref{eq:corr}, we recover the well-known formula
\begin{equation}
\label{eq:resplin_app}
 R_t=A^t=C_tC_0^{-1}\,,
\end{equation} 
valid for linear Markov systems at discrete times, where we have introduced the covariance matrix $C_t=\langle \vect{x}_t \vect{x}^{T}_0\rangle$. 
From Eq.~\eqref{eq:fgen} it is now clear that in these systems one can 
observe non-vanishing responses from $x^{(j)}$ to $x^{(k)}$, 
for any possible choice of $\mathcal{F}(x^{(k)})$, only if $R_t^{kj}\ne 0$; 
therefore the knowledge of the matrix $R_t$ (i.e., $\mathcal{F}(x)=x$) 
is sufficient to establish the causal links in a linear Markov 
dynamics.

\section{Sketch of the computation of Eq.~\eqref{eq:gpresult}}
\label{app:gp}

In this Appendix we sketch the computation to derive Eq.~\eqref{eq:gpresult}.
First, by multiplying Eq.~\eqref{eq:evo} by $\vect{x}_{t+1}^T$ and by $\vect{x}_{t}^T$ to the right, and taking averages on the stationary joint p.d.f., we get
\begin{equation}
\begin{cases}
 C_0=AC_0A^{T}+B^2\\ 
 C_1=AC_0\,.
\end{cases}
\end{equation}
For the simple model~\eqref{eq:simplemodel}, by solving the above system 
one finds
\begin{equation}
\label{eq:matrixc}
 C_0=2 D_1\begin{pmatrix}
      1+3r&r\\
      r&r
     \end{pmatrix}
     \quad
     C_1=\sqrt{2} D_1\begin{pmatrix}
     1+4r&2r\\
      2r&r
     \end{pmatrix}
\end{equation} 
where $r=D_2/D_1$.

Now, we have to compute the amplitudes of the noises $\Delta_1$ and 
$\Delta_2$ in the two alternative models~\eqref{eq:modelgp1} 
and~\eqref{eq:modelgp2}.  
$\Delta_1$ is given by a linear regression analysis: 
Eq.~\eqref{eq:modelgp1} yields
\begin{equation}
\begin{cases}
 \langle x^{(1)}_{t+1}x^{(1)}_{t}\rangle=\alpha_1 \langle (x^{(1)}_{t})^2\rangle\\
 \langle (x^{(1)}_{t+1})^2\rangle=\alpha_1^2 \langle (x^{(1)}_{t})^2\rangle+ \Delta_1 
\end{cases}
\end{equation}
i.e.
\begin{equation}
\begin{cases}
 C_1^{11}=\alpha_1 C_0^{11}\\
 C_1^{11}=\alpha_1^2 C_0^{11}+\Delta_1\,, 
\end{cases}
\end{equation} 
in which the coefficients of the matrices $C_0$ and $C_1$ are given by Eq.~\eqref{eq:matrixc}. Simple algebra leads to
\begin{equation}
 \Delta_1=D_1 \dfrac{1+4r+r^2}{1+3r}\,.
\end{equation} 
Instead, $\Delta_2$ is clearly equal to $D_1$, since 
a similar regression analysis on model~\eqref{eq:modelgp2} shows that 
the best AR-model coincides with the original dynamics. 
The above values of $\Delta_1$ and $\Delta_2$ lead 
to Eq.~\eqref{eq:gpresult}. 
\quad\\
\quad\\

\section{Sketch of the computation of Eq.~\eqref{eq:teresult}}
\label{app:te}
To compute the TE for model~\eqref{eq:simplemodel}, it is useful to 
write down explicitly the following quantities, bearing in mind that 
all p.d.f.s refer here to linear Gaussian processes:
\begin{subequations}
 \begin{equation}
 \begin{aligned}
   \ln[p(x^{(1)}_{t+1}&|x^{(1)}_{t},x^{(2)}_{t})]= - \dfrac{1}{2}\ln(2 \pi D_1)\\&-\dfrac{(x^{(1)}_{t+1}-x^{(1)}_{t}/\sqrt{2}-x^{(2)}_{t}/\sqrt{2})^2}{2D_1}
 \end{aligned}
  \end{equation}  
 \begin{equation}
\ln[p(x^{(1)}_t)]=-\frac{1}{2}\ln(2\pi C_0^{11})-\dfrac{(x_{t}^{(1)})^2}{2C_0^{11}}
  \end{equation} 
 \begin{equation}
\ln[p(x^{(1)}_{t+1},x^{(1)}_{t})]=-\frac{1}{2}\ln(4 \pi^2 |\Sigma|)-\frac{1}{2}\vect{v}\Sigma^{-1}\vect{v}^T
\end{equation} 
\end{subequations}
where, in the last equation, 
$\vect{v}=(x_{t+1},x_t)^T$, $\Sigma=\langle \vect{v}\vect{v}^T\rangle$, 
while $|\cdot|$ represents the determinant. Matrix $C_0$ is defined by Eq.~\eqref{eq:matrixc}.

The above quantities have to be averaged over the joint stationary p.d.f. $p(x^{(1)}_{t+1},x^{(1)}_{t},x^{(2)}_{t})$. Recalling Eq.~\eqref{eq:matrixc} we get
\begin{subequations}
 \begin{equation}
   \langle\ln[p(x^{(1)}_{t+1}|x^{(1)}_{t},x^{(2)}_{t})]\rangle= - \dfrac{1}{2}\ln(2 \pi D_1)\\-1/2
  \end{equation}  
 \begin{equation}
\langle\ln[p(x^{(1)}_t)]\rangle=-\frac{1}{2}\ln[2\pi D_1(2+6r)]-1/2
  \end{equation} 
 \begin{equation}
\langle\ln[p(x^{(1)}_{t+1},x^{(1)}_{t})]\rangle=-\frac{1}{2}\ln[4 \pi^2 D_1^2(2+8r+4r^2)]-1\,.
\end{equation} 
\end{subequations}

The result in Eq.~\eqref{eq:teresult} is then readily recovered by noticing that
\begin{equation}
 \begin{aligned}
  TE_{2\to 1}&=\left\langle \ln \dfrac{p(x^{(1)}_{t+1}|x^{(1)}_{t},x^{(2)}_{t})}{p(x^{(1)}_{t+1}|x^{(1)}_{t})}\right\rangle\\
  &=\left\langle \ln \dfrac{p(x^{(1)}_{t+1}|x^{(1)}_{t},x^{(2)}_{t})p(x^{(1)}_{t})}{p(x^{(1)}_{t+1},x^{(1)}_{t})}\right\rangle\\
  &=\frac{1}{2}\ln[D_1^2(2+8r+4r^2)]+\\&-\frac{1}{2}\ln(D_1)-\frac{1}{2}\ln[D_1(2+6r)]\,.
 \end{aligned}
\end{equation}

\bibliography{biblio}

\begin{thebibliography}{40}%
\makeatletter
\providecommand \@ifxundefined [1]{%
 \@ifx{#1\undefined}
}%
\providecommand \@ifnum [1]{%
 \ifnum #1\expandafter \@firstoftwo
 \else \expandafter \@secondoftwo
 \fi
}%
\providecommand \@ifx [1]{%
 \ifx #1\expandafter \@firstoftwo
 \else \expandafter \@secondoftwo
 \fi
}%
\providecommand \natexlab [1]{#1}%
\providecommand \enquote  [1]{``#1''}%
\providecommand \bibnamefont  [1]{#1}%
\providecommand \bibfnamefont [1]{#1}%
\providecommand \citenamefont [1]{#1}%
\providecommand \href@noop [0]{\@secondoftwo}%
\providecommand \href [0]{\begingroup \@sanitize@url \@href}%
\providecommand \@href[1]{\@@startlink{#1}\@@href}%
\providecommand \@@href[1]{\endgroup#1\@@endlink}%
\providecommand \@sanitize@url [0]{\catcode `\\12\catcode `\$12\catcode
  `\&12\catcode `\#12\catcode `\^12\catcode `\_12\catcode `\%12\relax}%
\providecommand \@@startlink[1]{}%
\providecommand \@@endlink[0]{}%
\providecommand \url  [0]{\begingroup\@sanitize@url \@url }%
\providecommand \@url [1]{\endgroup\@href {#1}{\urlprefix }}%
\providecommand \urlprefix  [0]{URL }%
\providecommand \Eprint [0]{\href }%
\providecommand \doibase [0]{http://dx.doi.org/}%
\providecommand \selectlanguage [0]{\@gobble}%
\providecommand \bibinfo  [0]{\@secondoftwo}%
\providecommand \bibfield  [0]{\@secondoftwo}%
\providecommand \translation [1]{[#1]}%
\providecommand \BibitemOpen [0]{}%
\providecommand \bibitemStop [0]{}%
\providecommand \bibitemNoStop [0]{.\EOS\space}%
\providecommand \EOS [0]{\spacefactor3000\relax}%
\providecommand \BibitemShut  [1]{\csname bibitem#1\endcsname}%
\let\auto@bib@innerbib\@empty
\bibitem [{\citenamefont {Hume}(2001)}]{Hume}%
  \BibitemOpen
  \bibfield  {author} {\bibinfo {author} {\bibfnamefont {D.}~\bibnamefont
  {Hume}},\ }\href@noop {} {\emph {\bibinfo {title} {A Treatise of Human
  Nature: Vol.1}}},\ edited by\ \bibinfo {editor} {\bibfnamefont
  {D.}~\bibnamefont {Norton}}\ and\ \bibinfo {editor} {\bibfnamefont
  {M.}~\bibnamefont {Norton}}\ (\bibinfo  {publisher} {Oxford University
  Press},\ \bibinfo {year} {2001})\BibitemShut {NoStop}%
\bibitem [{\citenamefont {Pearl}(2009)}]{PearlBook}%
  \BibitemOpen
  \bibfield  {author} {\bibinfo {author} {\bibfnamefont {J.}~\bibnamefont
  {Pearl}},\ }\href@noop {} {\emph {\bibinfo {title} {Causality}}}\ (\bibinfo
  {publisher} {Cambridge University Press},\ \bibinfo {year}
  {2009})\BibitemShut {NoStop}%
\bibitem [{\citenamefont {Aurell}\ and\ \citenamefont
  {Del~Ferraro}(2016)}]{Aurell2016causal}%
  \BibitemOpen
  \bibfield  {author} {\bibinfo {author} {\bibfnamefont {E.}~\bibnamefont
  {Aurell}}\ and\ \bibinfo {author} {\bibfnamefont {G.}~\bibnamefont
  {Del~Ferraro}},\ }\href {\doibase
  https://doi.org/10.1088/1742-6596/699/1/012002} {\bibfield  {journal}
  {\bibinfo  {journal} {J. Phys.: Conf. Ser.}\ }\textbf {\bibinfo {volume}
  {699}},\ \bibinfo {pages} {012002} (\bibinfo {year} {2016})}\BibitemShut
  {NoStop}%
\bibitem [{\citenamefont {Zeng}\ and\ \citenamefont {Aurell}(2020)}]{zeng20}%
  \BibitemOpen
  \bibfield  {author} {\bibinfo {author} {\bibfnamefont {H.-L.}\ \bibnamefont
  {Zeng}}\ and\ \bibinfo {author} {\bibfnamefont {E.}~\bibnamefont {Aurell}},\
  }\href@noop {} {\enquote {\bibinfo {title} {Inverse {Ising} techniques to
  infer underlying mechanisms from data},}\ } (\bibinfo {year} {2020}),\
  \Eprint {http://arxiv.org/abs/2002.05222} {arXiv:2002.05222 [stat.OT]}
  \BibitemShut {NoStop}%
\bibitem [{\citenamefont {Friedrich}\ \emph {et~al.}(2011)\citenamefont
  {Friedrich}, \citenamefont {Peinke}, \citenamefont {Sahimi},\ and\
  \citenamefont {Tabar}}]{friedrich11}%
  \BibitemOpen
  \bibfield  {author} {\bibinfo {author} {\bibfnamefont {R.}~\bibnamefont
  {Friedrich}}, \bibinfo {author} {\bibfnamefont {J.}~\bibnamefont {Peinke}},
  \bibinfo {author} {\bibfnamefont {M.}~\bibnamefont {Sahimi}}, \ and\ \bibinfo
  {author} {\bibfnamefont {M.~R.~R.}\ \bibnamefont {Tabar}},\ }\href {\doibase
  https://doi.org/10.1016/j.physrep.2011.05.003} {\bibfield  {journal}
  {\bibinfo  {journal} {Physics Reports}\ }\textbf {\bibinfo {volume} {506}},\
  \bibinfo {pages} {87 } (\bibinfo {year} {2011})}\BibitemShut {NoStop}%
\bibitem [{\citenamefont {Zeng}\ \emph {et~al.}(2011)\citenamefont {Zeng},
  \citenamefont {Aurell}, \citenamefont {Alava},\ and\ \citenamefont
  {Mahmoudi}}]{zeng11}%
  \BibitemOpen
  \bibfield  {author} {\bibinfo {author} {\bibfnamefont {H.-L.}\ \bibnamefont
  {Zeng}}, \bibinfo {author} {\bibfnamefont {E.}~\bibnamefont {Aurell}},
  \bibinfo {author} {\bibfnamefont {M.}~\bibnamefont {Alava}}, \ and\ \bibinfo
  {author} {\bibfnamefont {H.}~\bibnamefont {Mahmoudi}},\ }\href {\doibase
  10.1103/PhysRevE.83.041135} {\bibfield  {journal} {\bibinfo  {journal} {Phys.
  Rev. E}\ }\textbf {\bibinfo {volume} {83}},\ \bibinfo {pages} {041135}
  (\bibinfo {year} {2011})}\BibitemShut {NoStop}%
\bibitem [{\citenamefont {Baldovin}\ \emph {et~al.}(2018)\citenamefont
  {Baldovin}, \citenamefont {Cecconi}, \citenamefont {Cencini}, \citenamefont
  {Puglisi},\ and\ \citenamefont {Vulpiani}}]{baldovin18}%
  \BibitemOpen
  \bibfield  {author} {\bibinfo {author} {\bibfnamefont {M.}~\bibnamefont
  {Baldovin}}, \bibinfo {author} {\bibfnamefont {F.}~\bibnamefont {Cecconi}},
  \bibinfo {author} {\bibfnamefont {M.}~\bibnamefont {Cencini}}, \bibinfo
  {author} {\bibfnamefont {A.}~\bibnamefont {Puglisi}}, \ and\ \bibinfo
  {author} {\bibfnamefont {A.}~\bibnamefont {Vulpiani}},\ }\href {\doibase
  https://doi.org/10.3390/e20100807} {\bibfield  {journal} {\bibinfo  {journal}
  {Entropy}\ }\textbf {\bibinfo {volume} {20}},\ \bibinfo {pages} {807}
  (\bibinfo {year} {2018})}\BibitemShut {NoStop}%
\bibitem [{\citenamefont {Ferretti}\ \emph {et~al.}(2019)\citenamefont
  {Ferretti}, \citenamefont {Chardès}, \citenamefont {Mora}, \citenamefont
  {Walczak},\ and\ \citenamefont {Giardina}}]{ferretti19}%
  \BibitemOpen
  \bibfield  {author} {\bibinfo {author} {\bibfnamefont {F.}~\bibnamefont
  {Ferretti}}, \bibinfo {author} {\bibfnamefont {V.}~\bibnamefont {Chardès}},
  \bibinfo {author} {\bibfnamefont {T.}~\bibnamefont {Mora}}, \bibinfo {author}
  {\bibfnamefont {A.~M.}\ \bibnamefont {Walczak}}, \ and\ \bibinfo {author}
  {\bibfnamefont {I.}~\bibnamefont {Giardina}},\ }\href@noop {} {\enquote
  {\bibinfo {title} {Building general {Langevin} models from discrete data
  sets},}\ } (\bibinfo {year} {2019}),\ \Eprint
  {http://arxiv.org/abs/1912.10491} {arXiv:1912.10491 [q-bio.QM]} \BibitemShut
  {NoStop}%
\bibitem [{\citenamefont {Simon}(1954)}]{simon54}%
  \BibitemOpen
  \bibfield  {author} {\bibinfo {author} {\bibfnamefont {H.~A.}\ \bibnamefont
  {Simon}},\ }\href {\doibase 10.1080/01621459.1954.10483515} {\bibfield
  {journal} {\bibinfo  {journal} {Journal of the American Statistical
  Association}\ }\textbf {\bibinfo {volume} {49}},\ \bibinfo {pages} {467}
  (\bibinfo {year} {1954})}\BibitemShut {NoStop}%
\bibitem [{\citenamefont {Hlav{\'a}{\v{c}}kov{\'a}-Schindler}\ \emph
  {et~al.}(2007)\citenamefont {Hlav{\'a}{\v{c}}kov{\'a}-Schindler},
  \citenamefont {Palu{\v{s}}}, \citenamefont {Vejmelka},\ and\ \citenamefont
  {Bhattacharya}}]{CausaInfoTh}%
  \BibitemOpen
  \bibfield  {author} {\bibinfo {author} {\bibfnamefont {K.}~\bibnamefont
  {Hlav{\'a}{\v{c}}kov{\'a}-Schindler}}, \bibinfo {author} {\bibfnamefont
  {M.}~\bibnamefont {Palu{\v{s}}}}, \bibinfo {author} {\bibfnamefont
  {M.}~\bibnamefont {Vejmelka}}, \ and\ \bibinfo {author} {\bibfnamefont
  {J.}~\bibnamefont {Bhattacharya}},\ }\href@noop {} {\bibfield  {journal}
  {\bibinfo  {journal} {Phys. Rep.}\ }\textbf {\bibinfo {volume} {441}},\
  \bibinfo {pages} {1} (\bibinfo {year} {2007})}\BibitemShut {NoStop}%
\bibitem [{\citenamefont {Atmanspacher}\ and\ \citenamefont
  {Martin}(2019)}]{atmanspacher19}%
  \BibitemOpen
  \bibfield  {author} {\bibinfo {author} {\bibfnamefont {H.}~\bibnamefont
  {Atmanspacher}}\ and\ \bibinfo {author} {\bibfnamefont {M.}~\bibnamefont
  {Martin}},\ }\href {\doibase 10.3390/info10090272} {\bibfield  {journal}
  {\bibinfo  {journal} {Information}\ }\textbf {\bibinfo {volume} {10}},\
  \bibinfo {pages} {272} (\bibinfo {year} {2019})}\BibitemShut {NoStop}%
\bibitem [{\citenamefont {Granger}(1969)}]{Granger69}%
  \BibitemOpen
  \bibfield  {author} {\bibinfo {author} {\bibfnamefont {C.~W.}\ \bibnamefont
  {Granger}},\ }\href@noop {} {\bibfield  {journal} {\bibinfo  {journal}
  {Econometrica}\ }\textbf {\bibinfo {volume} {37}},\ \bibinfo {pages} {424}
  (\bibinfo {year} {1969})}\BibitemShut {NoStop}%
\bibitem [{\citenamefont {Bressler}\ and\ \citenamefont
  {Seth}(2011)}]{bressler11}%
  \BibitemOpen
  \bibfield  {author} {\bibinfo {author} {\bibfnamefont {S.~L.}\ \bibnamefont
  {Bressler}}\ and\ \bibinfo {author} {\bibfnamefont {A.~K.}\ \bibnamefont
  {Seth}},\ }\href@noop {} {\bibfield  {journal} {\bibinfo  {journal}
  {Neuroimage}\ }\textbf {\bibinfo {volume} {58}},\ \bibinfo {pages} {323}
  (\bibinfo {year} {2011})}\BibitemShut {NoStop}%
\bibitem [{\citenamefont {Barrett}\ \emph {et~al.}(2010)\citenamefont
  {Barrett}, \citenamefont {Barnett},\ and\ \citenamefont {Seth}}]{barrett10}%
  \BibitemOpen
  \bibfield  {author} {\bibinfo {author} {\bibfnamefont {A.~B.}\ \bibnamefont
  {Barrett}}, \bibinfo {author} {\bibfnamefont {L.}~\bibnamefont {Barnett}}, \
  and\ \bibinfo {author} {\bibfnamefont {A.~K.}\ \bibnamefont {Seth}},\ }\href
  {\doibase 10.1103/PhysRevE.81.041907} {\bibfield  {journal} {\bibinfo
  {journal} {Phys. Rev. E}\ }\textbf {\bibinfo {volume} {81}},\ \bibinfo
  {pages} {041907} (\bibinfo {year} {2010})}\BibitemShut {NoStop}%
\bibitem [{\citenamefont {Cadotte}\ \emph {et~al.}(2008)\citenamefont
  {Cadotte}, \citenamefont {DeMarse}, \citenamefont {He},\ and\ \citenamefont
  {Ding}}]{cadotte08}%
  \BibitemOpen
  \bibfield  {author} {\bibinfo {author} {\bibfnamefont {A.~J.}\ \bibnamefont
  {Cadotte}}, \bibinfo {author} {\bibfnamefont {T.~B.}\ \bibnamefont
  {DeMarse}}, \bibinfo {author} {\bibfnamefont {P.}~\bibnamefont {He}}, \ and\
  \bibinfo {author} {\bibfnamefont {M.}~\bibnamefont {Ding}},\ }\href {\doibase
  10.1371/journal.pone.0003355} {\bibfield  {journal} {\bibinfo  {journal}
  {PLOS ONE}\ }\textbf {\bibinfo {volume} {3}},\ \bibinfo {pages} {1} (\bibinfo
  {year} {2008})}\BibitemShut {NoStop}%
\bibitem [{\citenamefont {Schreiber}(2000)}]{shreiber00}%
  \BibitemOpen
  \bibfield  {author} {\bibinfo {author} {\bibfnamefont {T.}~\bibnamefont
  {Schreiber}},\ }\href {\doibase 10.1103/PhysRevLett.85.461} {\bibfield
  {journal} {\bibinfo  {journal} {Phys. Rev. Lett.}\ }\textbf {\bibinfo
  {volume} {85}},\ \bibinfo {pages} {461} (\bibinfo {year} {2000})}\BibitemShut
  {NoStop}%
\bibitem [{\citenamefont {Bossomaier}\ \emph {et~al.}(2016)\citenamefont
  {Bossomaier}, \citenamefont {Barnett}, \citenamefont {Harr{\'e}},\ and\
  \citenamefont {Lizier}}]{bossomaier}%
  \BibitemOpen
  \bibfield  {author} {\bibinfo {author} {\bibfnamefont {T.}~\bibnamefont
  {Bossomaier}}, \bibinfo {author} {\bibfnamefont {L.}~\bibnamefont {Barnett}},
  \bibinfo {author} {\bibfnamefont {M.}~\bibnamefont {Harr{\'e}}}, \ and\
  \bibinfo {author} {\bibfnamefont {J.~T.}\ \bibnamefont {Lizier}},\
  }\href@noop {} {\emph {\bibinfo {title} {An Introduction to Transfer
  Entropy}}}\ (\bibinfo  {publisher} {Springer},\ \bibinfo {year}
  {2016})\BibitemShut {NoStop}%
\bibitem [{\citenamefont {Runge}\ \emph
  {et~al.}(2012{\natexlab{a}})\citenamefont {Runge}, \citenamefont {Heitzig},
  \citenamefont {Petoukhov},\ and\ \citenamefont {Kurths}}]{runge12}%
  \BibitemOpen
  \bibfield  {author} {\bibinfo {author} {\bibfnamefont {J.}~\bibnamefont
  {Runge}}, \bibinfo {author} {\bibfnamefont {J.}~\bibnamefont {Heitzig}},
  \bibinfo {author} {\bibfnamefont {V.}~\bibnamefont {Petoukhov}}, \ and\
  \bibinfo {author} {\bibfnamefont {J.}~\bibnamefont {Kurths}},\ }\href
  {\doibase 10.1103/PhysRevLett.108.258701} {\bibfield  {journal} {\bibinfo
  {journal} {Phys. Rev. Lett.}\ }\textbf {\bibinfo {volume} {108}},\ \bibinfo
  {pages} {258701} (\bibinfo {year} {2012}{\natexlab{a}})}\BibitemShut
  {NoStop}%
\bibitem [{\citenamefont {Runge}\ \emph
  {et~al.}(2012{\natexlab{b}})\citenamefont {Runge}, \citenamefont {Heitzig},
  \citenamefont {Marwan},\ and\ \citenamefont {Kurths}}]{runge12pre}%
  \BibitemOpen
  \bibfield  {author} {\bibinfo {author} {\bibfnamefont {J.}~\bibnamefont
  {Runge}}, \bibinfo {author} {\bibfnamefont {J.}~\bibnamefont {Heitzig}},
  \bibinfo {author} {\bibfnamefont {N.}~\bibnamefont {Marwan}}, \ and\ \bibinfo
  {author} {\bibfnamefont {J.}~\bibnamefont {Kurths}},\ }\href {\doibase
  10.1103/PhysRevE.86.061121} {\bibfield  {journal} {\bibinfo  {journal} {Phys.
  Rev. E}\ }\textbf {\bibinfo {volume} {86}},\ \bibinfo {pages} {061121}
  (\bibinfo {year} {2012}{\natexlab{b}})}\BibitemShut {NoStop}%
\bibitem [{\citenamefont {Sun}\ \emph {et~al.}(2015)\citenamefont {Sun},
  \citenamefont {Taylor},\ and\ \citenamefont {Bollt}}]{sun15}%
  \BibitemOpen
  \bibfield  {author} {\bibinfo {author} {\bibfnamefont {J.}~\bibnamefont
  {Sun}}, \bibinfo {author} {\bibfnamefont {D.}~\bibnamefont {Taylor}}, \ and\
  \bibinfo {author} {\bibfnamefont {E.~M.}\ \bibnamefont {Bollt}},\ }\href
  {\doibase 10.1137/140956166} {\bibfield  {journal} {\bibinfo  {journal} {SIAM
  Journal on Applied Dynamical Systems}\ }\textbf {\bibinfo {volume} {14}},\
  \bibinfo {pages} {73} (\bibinfo {year} {2015})}\BibitemShut {NoStop}%
\bibitem [{\citenamefont {Ito}\ and\ \citenamefont {Sagawa}(2013)}]{ito13}%
  \BibitemOpen
  \bibfield  {author} {\bibinfo {author} {\bibfnamefont {S.}~\bibnamefont
  {Ito}}\ and\ \bibinfo {author} {\bibfnamefont {T.}~\bibnamefont {Sagawa}},\
  }\href {\doibase 10.1103/PhysRevLett.111.180603} {\bibfield  {journal}
  {\bibinfo  {journal} {Phys. Rev. Lett.}\ }\textbf {\bibinfo {volume} {111}},\
  \bibinfo {pages} {180603} (\bibinfo {year} {2013})}\BibitemShut {NoStop}%
\bibitem [{\citenamefont {Auconi}\ \emph {et~al.}(2019)\citenamefont {Auconi},
  \citenamefont {Giansanti},\ and\ \citenamefont {Klipp}}]{auconi19}%
  \BibitemOpen
  \bibfield  {author} {\bibinfo {author} {\bibfnamefont {A.}~\bibnamefont
  {Auconi}}, \bibinfo {author} {\bibfnamefont {A.}~\bibnamefont {Giansanti}}, \
  and\ \bibinfo {author} {\bibfnamefont {E.}~\bibnamefont {Klipp}},\
  }\href@noop {} {\bibfield  {journal} {\bibinfo  {journal} {Entropy}\ }\textbf
  {\bibinfo {volume} {21}},\ \bibinfo {pages} {177} (\bibinfo {year}
  {2019})}\BibitemShut {NoStop}%
\bibitem [{\citenamefont {Barnett}\ \emph {et~al.}(2009)\citenamefont
  {Barnett}, \citenamefont {Barrett},\ and\ \citenamefont {Seth}}]{barnett09}%
  \BibitemOpen
  \bibfield  {author} {\bibinfo {author} {\bibfnamefont {L.}~\bibnamefont
  {Barnett}}, \bibinfo {author} {\bibfnamefont {A.~B.}\ \bibnamefont
  {Barrett}}, \ and\ \bibinfo {author} {\bibfnamefont {A.~K.}\ \bibnamefont
  {Seth}},\ }\href {\doibase 10.1103/PhysRevLett.103.238701} {\bibfield
  {journal} {\bibinfo  {journal} {Phys. Rev. Lett.}\ }\textbf {\bibinfo
  {volume} {103}},\ \bibinfo {pages} {238701} (\bibinfo {year}
  {2009})}\BibitemShut {NoStop}%
\bibitem [{\citenamefont
  {Hlav{\'a}{\v{c}}kov{\'a}-Schindler}(2011)}]{Equivalence}%
  \BibitemOpen
  \bibfield  {author} {\bibinfo {author} {\bibfnamefont {K.}~\bibnamefont
  {Hlav{\'a}{\v{c}}kov{\'a}-Schindler}},\ }\href@noop {} {\bibfield  {journal}
  {\bibinfo  {journal} {Applied Mathematical Science}\ }\textbf {\bibinfo
  {volume} {5}},\ \bibinfo {pages} {3637} (\bibinfo {year} {2011})}\BibitemShut
  {NoStop}%
\bibitem [{\citenamefont {Barrett}\ and\ \citenamefont
  {Barnett}(2013)}]{barrett13}%
  \BibitemOpen
  \bibfield  {author} {\bibinfo {author} {\bibfnamefont {A.}~\bibnamefont
  {Barrett}}\ and\ \bibinfo {author} {\bibfnamefont {L.}~\bibnamefont
  {Barnett}},\ }\href {\doibase 10.3389/fninf.2013.00006} {\bibfield  {journal}
  {\bibinfo  {journal} {Frontiers in Neuroinformatics}\ }\textbf {\bibinfo
  {volume} {7}},\ \bibinfo {pages} {6} (\bibinfo {year} {2013})}\BibitemShut
  {NoStop}%
\bibitem [{\citenamefont {Barnett}\ \emph {et~al.}(2018)\citenamefont
  {Barnett}, \citenamefont {Barrett},\ and\ \citenamefont {Seth}}]{barnett18}%
  \BibitemOpen
  \bibfield  {author} {\bibinfo {author} {\bibfnamefont {L.}~\bibnamefont
  {Barnett}}, \bibinfo {author} {\bibfnamefont {A.~B.}\ \bibnamefont
  {Barrett}}, \ and\ \bibinfo {author} {\bibfnamefont {A.~K.}\ \bibnamefont
  {Seth}},\ }\href {\doibase 10.1073/pnas.1714497115} {\bibfield  {journal}
  {\bibinfo  {journal} {Proceedings of the National Academy of Sciences}\
  }\textbf {\bibinfo {volume} {115}},\ \bibinfo {pages} {E6676} (\bibinfo
  {year} {2018})}\BibitemShut {NoStop}%
\bibitem [{\citenamefont {Marini Bettolo~Marconi}\ \emph
  {et~al.}(2008)\citenamefont {Marini Bettolo~Marconi}, \citenamefont
  {Puglisi}, \citenamefont {Rondoni},\ and\ \citenamefont
  {Vulpiani}}]{marconi08}%
  \BibitemOpen
  \bibfield  {author} {\bibinfo {author} {\bibfnamefont {U.}~\bibnamefont
  {Marini Bettolo~Marconi}}, \bibinfo {author} {\bibfnamefont {A.}~\bibnamefont
  {Puglisi}}, \bibinfo {author} {\bibfnamefont {L.}~\bibnamefont {Rondoni}}, \
  and\ \bibinfo {author} {\bibfnamefont {A.}~\bibnamefont {Vulpiani}},\ }\href
  {\doibase https://doi.org/10.1016/j.physrep.2008.02.002} {\bibfield
  {journal} {\bibinfo  {journal} {Phys. Rep.}\ }\textbf {\bibinfo {volume}
  {461}},\ \bibinfo {pages} {111 } (\bibinfo {year} {2008})}\BibitemShut
  {NoStop}%
\bibitem [{\citenamefont {Ay}\ and\ \citenamefont {Polani}(2008)}]{ay08}%
  \BibitemOpen
  \bibfield  {author} {\bibinfo {author} {\bibfnamefont {N.}~\bibnamefont
  {Ay}}\ and\ \bibinfo {author} {\bibfnamefont {D.}~\bibnamefont {Polani}},\
  }\href {\doibase 10.1142/S0219525908001465} {\bibfield  {journal} {\bibinfo
  {journal} {Advances in Complex Systems}\ }\textbf {\bibinfo {volume} {11}},\
  \bibinfo {pages} {17} (\bibinfo {year} {2008})}\BibitemShut {NoStop}%
\bibitem [{\citenamefont {Kubo~R.}(1991)}]{kubo_response}%
  \BibitemOpen
  \bibfield  {author} {\bibinfo {author} {\bibfnamefont {H.~N.}\ \bibnamefont
  {Kubo~R.}, \bibfnamefont {Toda~M.}},\ }\enquote {\bibinfo {title}
  {Statistical mechanics of linear response},}\ in\ \href@noop {} {\emph
  {\bibinfo {booktitle} {Statistical Physics II}}}\ (\bibinfo  {publisher}
  {Springer, Berlin, Heidelberg},\ \bibinfo {year} {1991})\BibitemShut
  {NoStop}%
\bibitem [{\citenamefont {Livi}\ and\ \citenamefont
  {Politi}(2017)}]{LiviPoliti}%
  \BibitemOpen
  \bibfield  {author} {\bibinfo {author} {\bibfnamefont {R.}~\bibnamefont
  {Livi}}\ and\ \bibinfo {author} {\bibfnamefont {P.}~\bibnamefont {Politi}},\
  }\href@noop {} {\emph {\bibinfo {title} {Nonequilibrium Statistical Physics:
  A Modern Perspective}}}\ (\bibinfo  {publisher} {Cambridge University
  Press},\ \bibinfo {year} {2017})\BibitemShut {NoStop}%
\bibitem [{\citenamefont {Messerli}(2012)}]{messerli12}%
  \BibitemOpen
  \bibfield  {author} {\bibinfo {author} {\bibfnamefont {F.~H.}\ \bibnamefont
  {Messerli}},\ }\href {\doibase 10.1056/NEJMon1211064} {\bibfield  {journal}
  {\bibinfo  {journal} {N. Engl. J. Med.}\ }\textbf {\bibinfo {volume} {367}},\
  \bibinfo {pages} {1562} (\bibinfo {year} {2012})}\BibitemShut {NoStop}%
\bibitem [{\citenamefont {Paluš}(2007)}]{palus07}%
  \BibitemOpen
  \bibfield  {author} {\bibinfo {author} {\bibfnamefont {M.}~\bibnamefont
  {Paluš}},\ }\href {\doibase 10.1080/00107510801959206} {\bibfield  {journal}
  {\bibinfo  {journal} {Contemporary Physics}\ }\textbf {\bibinfo {volume}
  {48}},\ \bibinfo {pages} {307} (\bibinfo {year} {2007})}\BibitemShut
  {NoStop}%
\bibitem [{\citenamefont {Liang}(2016)}]{liang16}%
  \BibitemOpen
  \bibfield  {author} {\bibinfo {author} {\bibfnamefont {X.~S.}\ \bibnamefont
  {Liang}},\ }\href {\doibase 10.1103/PhysRevE.94.052201} {\bibfield  {journal}
  {\bibinfo  {journal} {Phys. Rev. E}\ }\textbf {\bibinfo {volume} {94}},\
  \bibinfo {pages} {052201} (\bibinfo {year} {2016})}\BibitemShut {NoStop}%
\bibitem [{\citenamefont {Tang}\ and\ \citenamefont {Kaneko}(2020)}]{Kaneko}%
  \BibitemOpen
  \bibfield  {author} {\bibinfo {author} {\bibfnamefont {Q.-Y.}\ \bibnamefont
  {Tang}}\ and\ \bibinfo {author} {\bibfnamefont {K.}~\bibnamefont {Kaneko}},\
  }\href@noop {} {\bibfield  {journal} {\bibinfo  {journal} {PLOS Comput.
  Biol.}\ }\textbf {\bibinfo {volume} {16}},\ \bibinfo {pages} {e1007670}
  (\bibinfo {year} {2020})}\BibitemShut {NoStop}%
\bibitem [{\citenamefont {Piazza}\ \emph {et~al.}(2009)\citenamefont {Piazza},
  \citenamefont {De~Los~Rios},\ and\ \citenamefont {Cecconi}}]{Piazza}%
  \BibitemOpen
  \bibfield  {author} {\bibinfo {author} {\bibfnamefont {F.}~\bibnamefont
  {Piazza}}, \bibinfo {author} {\bibfnamefont {P.}~\bibnamefont {De~Los~Rios}},
  \ and\ \bibinfo {author} {\bibfnamefont {F.}~\bibnamefont {Cecconi}},\ }\href
  {\doibase 10.1103/PhysRevLett.102.218104} {\bibfield  {journal} {\bibinfo
  {journal} {Phys. Rev. Lett.}\ }\textbf {\bibinfo {volume} {102}},\ \bibinfo
  {pages} {218104} (\bibinfo {year} {2009})}\BibitemShut {NoStop}%
\bibitem [{\citenamefont {Rümelin}(1982)}]{rumelin82}%
  \BibitemOpen
  \bibfield  {author} {\bibinfo {author} {\bibfnamefont {W.}~\bibnamefont
  {Rümelin}},\ }\href {http://www.jstor.org/stable/2156972} {\bibfield
  {journal} {\bibinfo  {journal} {SIAM Journal on Numerical Analysis}\ }\textbf
  {\bibinfo {volume} {19}},\ \bibinfo {pages} {604} (\bibinfo {year}
  {1982})}\BibitemShut {NoStop}%
\bibitem [{\citenamefont {Takens}(1981)}]{Takens81}%
  \BibitemOpen
  \bibfield  {author} {\bibinfo {author} {\bibfnamefont {F.}~\bibnamefont
  {Takens}},\ }in\ \href@noop {} {\emph {\bibinfo {booktitle} {Dynamical
  systems and turbulence}}},\ \bibinfo {series} {Lect. Notes in Mathematics},
  Vol.\ \bibinfo {volume} {898},\ \bibinfo {editor} {edited by\ \bibinfo
  {editor} {\bibfnamefont {D.~A.}\ \bibnamefont {Rand~and}}\ and\ \bibinfo
  {editor} {\bibfnamefont {L.~S.}\ \bibnamefont {Young}}}\ (\bibinfo
  {publisher} {Springer},\ \bibinfo {year} {1981})\ pp.\ \bibinfo {pages}
  {366--381}\BibitemShut {NoStop}%
\bibitem [{\citenamefont {Kantz}\ and\ \citenamefont
  {Schreiber}(2003)}]{kantz03}%
  \BibitemOpen
  \bibfield  {author} {\bibinfo {author} {\bibfnamefont {H.}~\bibnamefont
  {Kantz}}\ and\ \bibinfo {author} {\bibfnamefont {T.}~\bibnamefont
  {Schreiber}},\ }\href {\doibase 10.1017/CBO9780511755798} {\emph {\bibinfo
  {title} {Nonlinear Time Series Analysis}}},\ \bibinfo {edition} {2nd}\ ed.\
  (\bibinfo  {publisher} {Cambridge University Press},\ \bibinfo {year}
  {2003})\BibitemShut {NoStop}%
\bibitem [{\citenamefont {Onsager}\ and\ \citenamefont
  {Machlup}(1953)}]{onsager53}%
  \BibitemOpen
  \bibfield  {author} {\bibinfo {author} {\bibfnamefont {L.}~\bibnamefont
  {Onsager}}\ and\ \bibinfo {author} {\bibfnamefont {S.}~\bibnamefont
  {Machlup}},\ }\href {\doibase 10.1103/PhysRev.91.1505} {\bibfield  {journal}
  {\bibinfo  {journal} {Phys. Rev.}\ }\textbf {\bibinfo {volume} {91}},\
  \bibinfo {pages} {1505} (\bibinfo {year} {1953})}\BibitemShut {NoStop}%
\bibitem [{\citenamefont {Ma}(1985)}]{ma85}%
  \BibitemOpen
  \bibfield  {author} {\bibinfo {author} {\bibfnamefont {S.-K.}\ \bibnamefont
  {Ma}},\ }\href@noop {} {\emph {\bibinfo {title} {Statistical Mechanics}}}\
  (\bibinfo  {publisher} {World Scientific},\ \bibinfo {address} {Phyladelphia
  Singapore},\ \bibinfo {year} {1985})\BibitemShut {NoStop}%
\end{thebibliography}%

\end{document}